\documentclass[twocolumn, numberedappendix, appendixfloats, twocolappendix, tighten]{openjournal}
\usepackage{natbib}
\usepackage{graphicx,amsmath,amssymb,amstext}
\usepackage{amsbsy,amsfonts,amsthm,color}
\usepackage{lipsum}
\usepackage[colorlinks,linkcolor=blue,citecolor=blue,urlcolor=blue ]{hyperref}
\usepackage[utf8]{inputenc}
\usepackage{float}
\usepackage[caption=false]{subfig}
\usepackage{upgreek}
\usepackage{multirow}
\usepackage{makecell}
\usepackage{orcidlink}
\usepackage{xspace}

\makeatletter
\xspaceaddexceptions{[]}
\makeatother

\definecolor{green}{rgb}{0,0.8,0}  

\newcommand{\sqdeg}{deg$^2$\xspace}
\newcommand{\um}{$\mu$m\xspace}
\newcommand{\ujy}{$\mu$Jy\xspace}
\newcommand{\mjysr}{MJy~sr$^\text{-1}$\xspace}
\newcommand{\Ihundred}{$I_\text{100}$\xspace}

\newcommand{\jwst}{\textit{JWST}\xspace}
\newcommand{\euclid}{\textit{Euclid}\xspace}
\newcommand{\herschel}{\textit{Herschel}\xspace}
\newcommand{\spitzer}{\textit{Spitzer}\xspace}

\newcommand{\conf}{$\sigma_{\text{conf}}$}
\newcommand{\inst}{$\sigma_{\text{inst}}$}
\newcommand{\total}{$\sigma_{\text{total}}$}
\newcommand{\xid}{\texttt{XID+}\xspace}
\newcommand{\stepwise}{\texttt{XID+stepwise}\xspace}

\newcommand\blfootnote[1]{%
  \begingroup
  \renewcommand\thefootnote{}\footnote{#1}%
  \addtocounter{footnote}{-1}%
  \endgroup
}

\newcommand{\valasym}[3]{\makebox[7.4em][l]{\makebox[3.6em][r]{#1}$^{+#2}_{-#3}$}}

\defcitealias{Donnellan24}{D24}
\defcitealias{Bethermin24}{B24}

\begin{document}

\title{XID+PRIMA, II: Stepping through Hyperspectral Imaging to deblend PRIMAger Beyond the Extragalactic Confusion Limit}

\author{J. M. S. Donnellan$^{1 \star \dagger}$\orcidlink{0000-0003-3331-0219}}
\author{B. Pautasso$^{1 \star}$\orcidlink{0009-0003-1839-591X}}
\author{S. J. Oliver$^{1}$\orcidlink{0000-0001-7862-1032}}
\author{M. B\'ethermin$^{2}$\orcidlink{0000-0002-3915-2015}}
\author{L. Bing$^{1}$\orcidlink{0000-0002-0440-7129}}
\author{A. Bolatto$^{3}$\orcidlink{0000-0002-5480-5686}}
\author{L. Ciesla$^{4}$\orcidlink{0000-0003-0541-2891}}
\author{W. Jellema$^{5,6}$\orcidlink{0000-0002-4837-9553}}
\author{D. Koopmans$^{5,6}$\orcidlink{0009-0009-7490-1526}}
\author{A. Pope$^{7}$\orcidlink{0000-0001-8592-2706}}
\author{S. Serjeant$^{8}$\orcidlink{0000-0002-0517-7943}}
\author{J.-D. T. Smith$^{9}$\orcidlink{0000-0003-1545-5078}}
\author{L. Wang$^{5,6}$\orcidlink{0000-0002-6736-9158}}

\blfootnote{$^{\star}$ Denotes principal authors with equal contribution.}
\blfootnote{$^{\dagger}$ Email: \href{mailto:j.donnellan@sussex.ac.uk}{j.donnellan@sussex.ac.uk}}

\affiliation{$^{1}$Astronomy Centre, University of Sussex, Falmer, Brighton BN1 9QH, UK}
\affiliation{$^{2}$Universit\'e de Strasbourg, CNRS, Observatoire astronomique de Strasbourg, UMR 7550, 67000 Strasbourg, France}
\affiliation{$^{3}$Department of Astronomy, University of Maryland, College Park, MD 20742, USA}
\affiliation{$^{4}$Aix Marseille Univ, CNRS, CNES, LAM, Marseille, France}
\affiliation{$^{5}$SRON Netherlands Institute for Space Research, Landleven 12, 9747 AD Groningen, The Netherlands}
\affiliation{$^{6}$Kapteyn Astronomical Institute, University of Groningen, Postbus 800, 9700 AV Groningen, The Netherlands}
\affiliation{$^{7}$Department of Astronomy, University of Massachusetts, Amherst, MA 01003, USA}
\affiliation{$^{8}$School of Physical Sciences, The Open University, Walton Hall, Milton Keynes, MK7 6AA, UK}
\affiliation{$^{9}$Ritter Astrophysical Research Center, Department of Physics \& Astronomy, University of Toledo, Toledo, OH 43606, USA}

\begin{abstract}

The PRobe far-Infrared Mission for Astrophysics (PRIMA) concept aims to map large areas with spectral coverage and sensitivities inaccessible to previous FIR space telescopes, covering 25--235~\um.
We synthesise images representing a deep imaging survey, with realistic instrumental and confusion noise, reflecting the latest PRIMAger instrument specifications.
We present a new Bayesian modelling approach \stepwise that exploits PRIMAger’s hyperspectral imaging to derive self-consistent, informative flux priors by sequentially propagating constraints from short to long wavelengths.
With \euclid-like prior source positions, this method recovers fluxes to within 20\% down to 0.2--0.7~mJy across 45--84~\um, which correspond to factors of 1.3--3.5 fainter than the confusion limit.
For the most confusion-dominated channels, accurate fluxes are measured down to 0.8, 2.7, 9.3, and 23.4~mJy at 92, 126, 183, and 235~\um, respectively, which are factors of 2--5 better than the confusion limit.
Using a deeper \euclid-based prior catalogue and weak ancillary flux priors at 25~\um yields further improvements, reaching up to a factor $\sim$5 fainter than the confusion limit at 96~\um.
Additionally, we demonstrate that positional priors from blind source detection followed by deblending via \xid enables PRIMAger to achieve sensitivity beyond the confusion limits using PRIMAger data alone.
We show that IR-luminous galaxies at $z \sim 2$ are robustly detected in a large fraction of the PRIMAger channels ($>$98\% in 12 out of the 16 considered channels), providing dense sampling of the FIR SED even for sources several factors below the classical confusion limit.
We explore the impact on our results for a range of systematic effects, including cirrus contamination, optical degradation, and calibration uncertainties.
These findings indicate that confusion noise will not limit the key science from PRIMA extragalactic imaging surveys when employing \xid.
\end{abstract}


\section{Introduction}
\label{sec:intro}

Observing galaxies at far-infrared (FIR) wavelengths is crucial to understanding the physical processes that govern their evolution.
Dust within galaxies is a key component of the interstellar medium \citep[ISM;][]{Draine2011} and can reprocess a significant fraction of the ultraviolet (UV) emission from hot massive stars, which trace star formation, into the FIR \citep{Madau2014}.
As a result, the peak of a galaxy's spectral energy density (SED) often falls within the FIR wavelength range ($\sim$100~\um rest-frame).
From this reprocessing by dust, roughly half of the total extragalactic background light originates from the infrared emission we receive from galaxies, known as the cosmic infrared background \citep[CIB;][]{Puget1996}, despite dust only making up a fraction of the total baryonic mass of a galaxy \citep{Hauser2001,Dole2006}.
Since the FIR emission is proportional to the obscured star formation rate \citep[SFR;][]{Kennicutt1998}, this leads to a substantial fraction of the total SFR density of the universe being obscured by dust, particularly during the peak epoch when most stars were formed \citep[cosmic noon: $z \sim 2$;][]{Dunlop2017, Zavala2021}.

Galactic dust also significantly changes how we observe galaxies through the attenuation of rest-frame UV to near-infrared (NIR) emission, whereby dust absorbs and scatters light into and out of the line of sight.
It is crucial to correct for this in order to avoid biases in the determination of physical properties of galaxies, such as the stellar mass, SFR, age, and metallicity \citep{Cardelli1989,Calzetti2000}.
Different star-dust geometries, dust grain types, and sizes can also significantly alter the attenuation (see \cite{Samil&Narayanan2020} for a review).
As such, comprehensive sampling of a galaxy's SED from the UV to FIR is required in order to better constrain this attenuation, particularly when applying SED-fitting techniques to determine galaxy properties (see reviews by \citealt{Walcher2011, Conroy2013}).
Moreover, dense sampling across the peak of the SED is required to accurately determine IR luminosity (LIR), dust mass, and dust temperature \citep{Casey2012,Farrah2025}.
Good coverage of the mid- to far-infrared (MIR-FIR) range captures and constrains emissions from polycyclic aromatic hydrocarbons (PAHs) and warm dust \citep[see reviews by][]{Casey2014,Li2020}, whilst also significantly helping to discriminate between contributions to a galaxy's SED from star formation and active galactic nuclei \citep[AGN;][]{Pope2008,Mullaney2011}.

Currently, the \textit{James Webb Space Telescope} \citep[\jwst;][]{JWST_Overview} probes the rest-frame UV-MIR and MIRI \citep{2023PASP..135d8003W} has been used to begin constraining PAH evolution and dust luminosities of galaxies out to $z \sim 2$ \citep[e.g.][]{Shivaei2024}.
However, \jwst is limited to relatively small area surveys ($< 0.4$~sq.~deg. surveyed with MIRI in deep extragalactic fields in the first 4 years of \jwst) and can only observe up to $\sim$25~\um.
The Atacama Large Millimetre Array \citep[ALMA;][]{ALMA_Overview}, on the other hand, primarily probes the sub-millimetre regime ($\gtrsim$ 300~\um) and can detect dust-obscured star-forming galaxies out to the highest redshifts \citep{Fudamoto2020,Gruppioni2020,Inami2022}, but is also limited in its small field of view.
Previous space-based FIR observatories such as the \textit{Herschel Space Observatory} \citep[\herschel;][]{Herschel_Overview} and \textit{Spitzer Space Observatory} \citep[\spitzer;][]{Spitzer_Overview} filled the wavelength coverage gap between \jwst and ALMA before ending observations over a decade ago, but these were limited in both spectral sampling capabilities and sensitivity.

To bridge the current sensitivity and wavelength gap, the PRobe Far-Infrared Mission for Astrophysics \citep[PRIMA;][]{Glenn2025} has been proposed.
PRIMA is a 1.8~m space-based telescope which will be cryogenically-cooled to 4.5~K and has two planned instruments: the Far-Infrared Enhanced Survey Spectrometer \citep[FIRESS;][]{Bradford2025} and the PRIMA Imager \citep[PRIMAger;][]{Ciesla2025}.
FIRESS provides continuous spectral coverage from 24 to 235~\um in two spectral resolution modes ($R \geq 85$ and $R~=~4400(112~\text{\um}/\lambda)$).
PRIMAger will perform imaging via two modes, with the first providing hyperspectral imaging with $R \sim 8$ linear variable filters (LVFs) from 24 to 84~\um.
The second provides both intensity and polarimetric imaging via four broadband filters from 92 to 235~\um. 

Both FIRESS and PRIMAger will provide significant improvements in sensitivity, mapping speed, and wavelength coverage compared to previous FIR observatories.
These will allow PRIMA to conduct large area extragalactic surveys (tens of sq.~deg.), comprehensively sampling the MIR-FIR SED of a large sample of galaxies out to $z \sim 4$ \citep{Burgarella2025, Boquien2025}.
In particular, it will enable the study of PAH emission \citep{Yoon2025}, constrain the dust-mass function \citep{Traina2025}, reveal dust-obscured AGN \citep{Donnan2025,Barchiesi2025}, and shed light on the co-evolution of supermassive black holes (SMBHs) and their host galaxies \citep{Faisst2025,Fernandez2025}.
Many other applications exploiting the capabilities of PRIMA have also been identified by the wider astronomy community \citep{Moullet2023, Moullet2025}.

In order to realise the full benefit from these gains in sensitivity in extragalactic imaging surveys with PRIMAger, it will be essential to reduce the impact of confusion noise.
Confusion noise limits the depth reached in FIR imaging data for space-based, single-dish observatories due to the limited mirror size \citep{Condon74}.
For a given mirror size, the angular resolution of the telescope decreases as wavelength increases, leading to the blurring of sources when the telescope beam is large compared to the average separation of those sources.
Confusion noise, for example, limited \herschel surveys to detect only the most extreme sources at high redshift \citep{Nguyen2010}. 

\cite{Bethermin24} (hereafter referred to as \citetalias{Bethermin24}) demonstrated that PRIMAger will be confusion-noise limited at wavelengths of $\lambda \gtrsim $ 45~\um for its deep extragalactic surveys.
However, \citet[][\citetalias{Donnellan24}]{Donnellan24} demonstrated that accurate flux measurements could be obtained below the confusion limits of simulated PRIMAger maps.
This was achieved by employing \xid \citep{Hurley}, a deblending tool which uses a probabilistic Bayesian framework which includes prior information of galaxy positions and fluxes to obtain the full posterior probability distributions of fluxes for those galaxies.
It has previously been used to perform source extraction for the \herschel Extragalactic Legacy Project \citep[HELP;][]{Shirley2019,Shirley2021}.

\cite{Hurley} demonstrated that \xid performs better in terms of flux accuracy and flux uncertainty accuracy for simulated \textit{Herschel}/SPIRE maps than previous prior based source extraction tools, such as \texttt{DESPHOT} \citep{desphot1, desphot2} and \texttt{LAMBDAR} \citep{lambdar}. Specifically, they found that \xid was a factor of 3 better than \texttt{DESPHOT} in terms of flux accuracy at 250~\um.
Other photometry tools such as \texttt{SExtractor} \citep{sextractor} or \texttt{DAOphot} \citep{daophot}, which assume sources are isolated, would perform worse than the above. They would also likely fail to accurately capture the correlated uncertainty between the extracted fluxes of blended sources in confusion-limited maps.
The full probabilistic Bayesian modelling of \xid does, however, allow for the recovery of the full posterior distributions of simultaneously fitted source parameters as well as explicitly modelling confusion noise.
Such modelling considerations are crucial for obtaining reliable photometry below the confusion limit, despite the higher computational cost.

Additional works showed that by applying prior flux information, e.g. from SED-fitting of ancillary photometry, further gains in flux accuracy and depth could be achieved \citep{Pearson2017, Pearson2018, Wang2021, Wang24}.
\citetalias{Donnellan24} found positional catalogues of high completeness coupled with weak prior flux information provided up to an order of magnitude improvement in source flux recovery relative to the confusion limits of simulated PRIMAger maps.
However, the prior flux information used in \citetalias{Donnellan24} was idealised, taking the true values of the simulation using a toy model, along with a physically unmotivated positional prior source catalogue. 

In this work, we set out to build upon the work of \citetalias{Donnellan24} and fully utilise the hyperspectral capabilities of PRIMAger to demonstrate that accurate prior flux knowledge can be obtained from the PRIMAger data itself.
We show that these informative flux priors can be used to deblend confusion-limited PRIMAger maps starting with realistic \euclid-based prior source catalogues to obtain accurate flux measurements for galaxies well below the confusion limit.
Additionally, we update our simulations to reflect the latest PRIMAger specifications \citep{Ciesla2025}, and investigate the impact of various systematics, such as galactic cirrus and calibration effects, on our results.

This paper is structured as follows.
In Section~\ref{sec:simulations}, we outline how the simulated PRIMAger maps are generated and highlight the sources of various systematics.
Section~\ref{sec:xid} describes the construction of positional prior source catalogues and the methodology of obtaining prior flux information.
Our main results are presented in Section~\ref{sec:results} along with the impact of systematics.
We then discuss the implications of our results in Section~\ref{sec:discussion} and make final conclusions in Section~\ref{sec:conclusion}.
\section{Simulations}
\label{sec:simulations}
To evaluate the flux-recovery performance of PRIMAger in extragalactic observations, we generate model FIR observations.
The model observations are composed of three components: the extragalactic sky (Section~\ref{sec:extragalactic_sky}), foreground emissions (Section~\ref{sec:foreground_emissions}), and the characteristics of PRIMAger itself (Section~\ref{sec:primager}).
These components are combined to produce a semi-realistic observation, on which we perform flux recovery tests.

\subsection{Extragalactic Sky}
\label{sec:extragalactic_sky}
We utilise the Simulated Infrared Dusty Extragalactic Sky \citep[SIDES\footnote{\url{https://data.lam.fr/sides/home}};][]{SIDES17, SIDES22, SIDES23} to model the FIR extragalactic sky.
We refer interested readers to these papers for details, but provide a brief summary here.

Originally presented in \cite{SIDES17}, SIDES is a semi-empirical simulation of the FIR and millimetre extragalactic sky.
Built upon the Bolshoi-Planck dark matter simulation \citep{Bolshoi1, Bolshoi2}, it populates dark matter haloes with galaxies of a given stellar mass using an abundance-matching technique.
Empirical scaling relations are used to assign galaxy properties such as SFR and LIR, calibrated to match observational constraints such as the evolving star-forming main-sequence and associated scatter \citep[e.g.][]{2015A&A...575A..74S}.
The simulation successfully reproduces a large set of observations including the source redshift distributions and the large-scale anisotropies of the CIB \citep{2013ApJ...772...77V}.
Crucially, SIDES is able to reproduce the FIR source number counts at various angular resolutions \citep{SIDES17, 2023A&A...677A..66B} as well as the distribution of pixel intensities of \herschel/SPIRE maps \citep{Glenn2010}, ensuring the inclusion of realistic confusion noise within any mock observation.

The resulting light-cone catalogue contains 5.6 million galaxies over an area of 1.96~\sqdeg spanning $0 < z < 10$.
\cite{SIDES23} applied the same methodology to the Uchuu simulation \citep{Uchuu21} to construct a larger 117~\sqdeg light-cone catalogue covering ${0 < z < 7}$.
We select a 2~\sqdeg region from this latter catalogue for our work.
Although we do not yet make full use of the large area of the updated SIDES catalogue, its large area enables the study of cosmic variance \citep[e.g.][references therein]{SIDES23} and the training of machine learning models \citep[e.g.][Koopmans et al. submitted]{Lauritsen21, Koopmans25} in the future.

\subsection{Foreground emissions}
\label{sec:foreground_emissions}
Observations of extragalactic sources in the FIR are contaminated by two components of foreground emission: zodiacal dust and galactic cirrus \citep{Low84}.

Zodiacal dust emission arises from the thermal re-radiation of absorbed solar radiation by warm interplanetary dust, with its spectrum peaking in the MIR \citep[$\sim$10~\um;][]{Leinert98, Hauser98}.
Although we acknowledge the impact it would have on our observations, particularly in the bluer PRIMAger channels, we do not model it in this work.
We justify its omission since it is smooth on the scale of point sources, and it should be possible to accurately remove its contribution using established zodiacal light models \citep[e.g.][]{Kelsall98}.

Galactic cirrus emission is instead produced by the thermal re-radiation of the energy flux from the ambient interstellar radiation field by much cooler interstellar dust.
Due to the lower temperature of the dust grains, its emission peaks in the FIR \citep[$\sim$150~\um;][]{Draine03, Zubko04}.

\subsubsection{Modelling cirrus}
\label{sec:modelling_cirrus}
We utilise the IRIS \citep{IRIS} 100~\um maps to predict the cirrus intensity (\Ihundred) in several proposed PRIMA deep fields.
After correcting the maps for residual zodiacal light and CIB contamination following \cite{MD07}, we find the considered fields lie in regions of low cirrus, with average intensities ranging from 0.4 to 2.6~\mjysr.
To cover a range of potential cirrus conditions, we model three cirrus intensities: 0.5, 1.5, and 2.5~\mjysr.

We adopt Fractional Brownian motion (fBm) as a common, if crude, method used to simulate cirrus emissions \citep[e.g.][references therein]{MD07}.
Although capable of replicating the self-similar nature of cirrus, it fails at producing the large-scale filamentary structures observed in real cirrus since the phase distribution of the fBm map is random.
However, since our analysis is principally concerned with fluctuations on or around the scale of point sources, and not those on larger scales, we adopt a fBm model as a suitable clumpy background to represent the effects of cirrus.
Further discussion of the fBm methodology is provided in Appendix~\ref{sec:appx_cirrus}

The structure map produced by the fBm process is generated in arbitrary units, normalised between 0 and 1.
To construct an intensity map, we scale the structure map such that variance at the scale of a point source matches the \cite{Gautier92} prediction, applying the corrections suggested in \cite{MD07} for low intensity (\Ihundred $< 10$~\mjysr) regions.
We additionally scale the intensity to the appropriate wavelength using the interstellar dust SED from \cite{Zubko04}.
\\
\subsection{PRIMAger}
\label{sec:primager}
As mentioned previously, PRIMAger will operate in two imaging modes.
The PRIMAger Hyperspectral Imager (PHI) uses LVFs to conduct hyperspectral imaging between 24 and 84~\um at a spectral resolution of $R = \lambda/\Delta\lambda \geq 8$, divided into two bands (PHI1 and PHI2), whilst the PRIMAger Polarimetric Imager (PPI) provides intensity and polarimetric imaging across four bands with $R \approx 4$, covering 92 to 235~\um \citep{Ciesla2025}.

\begin{deluxetable*}{lcccccc}
\tablecaption{PRIMAger channel properties and detection limits}

\tablehead{
\colhead{Channel} &
\colhead{\begin{tabular}{c}Peak\\Wavelength\\($\lambda_\text{peak}$)\end{tabular}} &
\colhead{\begin{tabular}{c}Filter\\FWHM\end{tabular}} &
\colhead{\begin{tabular}{c}Spectral\\Resolution\\($R$)\end{tabular}} &
\colhead{\begin{tabular}{c}Beam\\FWHM\end{tabular}} &
\colhead{\begin{tabular}{c}Point Source\\Sensitivity\\(5\inst)\end{tabular}} &
\colhead{\begin{tabular}{c}Confusion\\Limit\\(5\conf)\end{tabular}}
\\
\colhead{} &
\colhead{{[\um]}} &
\colhead{{[\um]}} &
\colhead{} &
\colhead{{[\arcsec]}} &
\colhead{{[\ujy]}} &
\colhead{{[\ujy]}}
}

\startdata
PHI1\_1 & 25.0 & 3.6  & 7.1  &  3.8 &  100 &     45 \\
PHI1\_2 & 27.8 & 4.2  & 6.7  &  4.0 &  111 &     50 \\
PHI1\_3 & 31.1 & 4.5  & 7.1  &  4.3 &  124 &     61 \\
PHI1\_4 & 34.3 & 4.9  & 7.1  &  4.6 &  137 &     75 \\
PHI1\_5 & 37.9 & 5.4  & 7.1  &  4.9 &  151 &     98 \\
PHI1\_6 & 42.3 & 6.4  & 6.7  &  5.3 &  169 &    134 \\
PHI2\_1 & 47.3 & 6.8  & 7.1  &  6.9 &  189 &    300 \\
PHI2\_2 & 52.1 & 7.5  & 7.1  &  7.2 &  208 &    414 \\
PHI2\_3 & 58.2 & 8.8  & 6.7  &  7.8 &  233 &    633 \\
PHI2\_4 & 65.0 & 9.3  & 7.1  &  8.4 &  260 &    967 \\
PHI2\_5 & 71.8 & 10.3 & 7.1  &  9.1 &  287 &  1,450 \\
PHI2\_6 & 80.0 & 12.1 & 6.7  &  9.9 &  320 &  2,310 \\
PPI1    & 92.0 & 23   & 4.0  & 10.8 &  145 &  4,090 \\
PPI2    & 126  & 32   & 3.9  & 14.8 &  209 & 12,900 \\
PPI3    & 183  & 46   & 4.0  & 21.4 &  294 & 32,100 \\
PPI4    & 235  & 58   & 4.1  & 27.5 &  375 & 45,200
\enddata

\tablecomments{
Details of the 12 coadded PRIMAger channels for the two hyperspectral bands, PHI1 and PHI2, and the four polarimetry broadband channels, PPI1-PPI4.
Peak wavelength and filter FWHMs are given for each channel.
For PPI, where the filters are modelled as tophats, we instead cite their central wavelength and total range respectively.
Spectral resolution of each channel is defined as $R = \lambda_\text{peak}/\text{FWHM}$.
Beam FWHMs are estimated for the baseline telescope aperture (1.8~m) and detector and pixel layout.
The point source sensitivities are given for a deep survey observed for $\sim$1500~hr~deg$^{-2}$ in the absence of confusion.
The classical confusion limits are estimated following \citetalias{Bethermin24}.
}
\label{tab:filter_stats}

\end{deluxetable*}

\subsubsection{PHI}
\label{sec:phi}
The two PHI bands span 24--45 and 45--84~\um respectively.
In \citetalias{Donnellan24}, each PHI band was approximated by 6 broad channels with $R = 10$, each consisting of a tophat filter and a Gaussian beam.

\begin{figure}
    \centering
    \includegraphics[width=\linewidth]{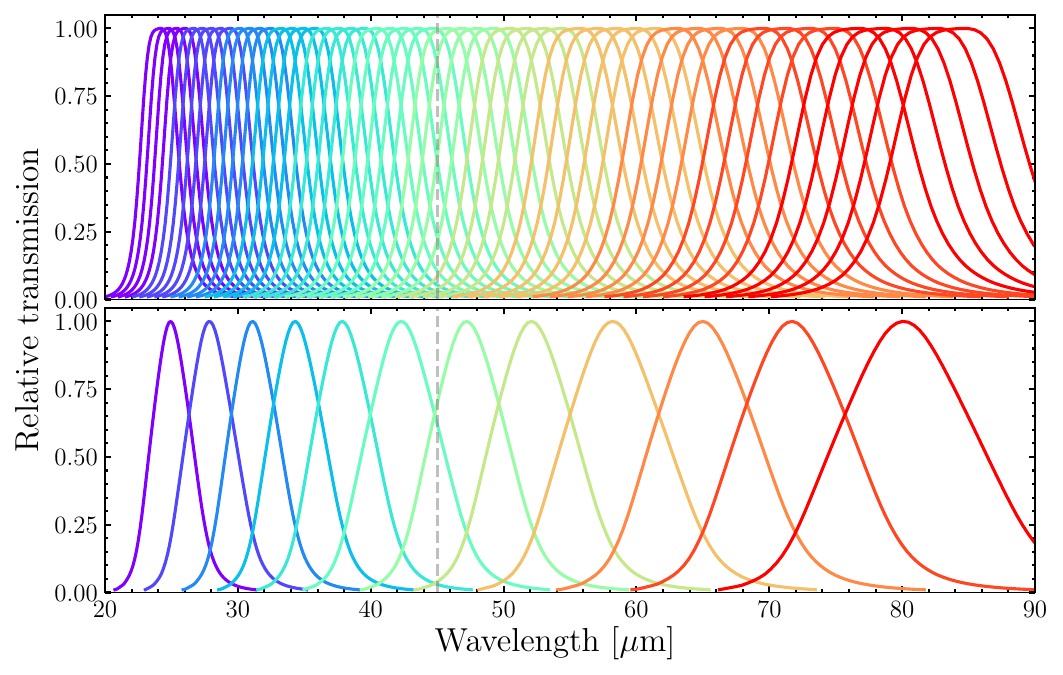}
    \caption{
    Filter transmission curves for \textbf{top:} all 52 PHI filters modelled in this work and \textbf{bottom:} the 12 coadded channels.
    Filters are coloured to match their associated channel.
    }
    \label{fig:filter_curves}
\end{figure}

With the development of the LVF, more realistic simulations of the filter transmissions than were used in \citetalias{Donnellan24} are now available \citep[][Jellema et al. in prep]{Jellema2024}.
The achieved profile is the result of a compromise between high spectral resolution and low out-of-band power.
In this work, we assume 52 logarithmically-sampled simulated filter profiles (26 per band), each with a spectral resolution of $R = 8$ and a profile aimed to be as close as possible to a tophat (Figure~\ref{fig:filter_curves}).
Further developments of the modelling of PHI spectral sampling and filter profile will be provided in Ciesla et al. (in prep).

More realistic optical modelling of the telescope beams has now been carried out.
Each filter now has a corresponding Airy disc beam that has been modified to include the effects of both central and tripod obscuration, as well as the coupling of the optics with the detectors (Jellema et al. in prep).

To improve signal-to-noise and reduce computational cost, we coadd nearby filters to form 12 channels comparable to those used in \citetalias{Donnellan24}.
This procedure is illustrated in Figure~\ref{fig:filter_curves}.

\subsubsection{PPI}
\label{sec:ppi}
In \citetalias{Donnellan24}, the four PPI bands were modelled with tophat filters and Gaussian beams.
In this work, the filters remain tophats, although PPI1 and PPI3 have been shifted to avoid prominent emission lines.
As with PHI, the beams are now modelled as Airy disc beams to more accurately reflect the telescope's optics.
Figure~\ref{fig:beams_2d} shows the 2D and radial profiles for three representative PRIMAger channels, PHI1\_1, PHI2\_1, and PPI1.

A summary of the 16 PRIMAger channels used in this work (12 for PHI and four for PPI) is given in Table~\ref{tab:filter_stats}.

\subsection{Observations and Mapmaking}
\label{sec:observations_mapmaking}
We convert the light-cone catalogue into sky maps using the SIDES Python mapmaker \citep{SIDES22}.
To summarise, the flux of each source is placed at the centre of its pixel, and then convolved with the telescope beams described previously. The resulting map is in units of Jy~beam$^{-1}$.
This process is repeated for each channel, using map pixels of 0.8, 1.3, and 2\farcs3 for PHI1, PHI2, and PPI respectively.

The resulting maps include confusion noise, which arises from the convolution of sources with the telescope beam \citep{Condon74}, but not instrumental noise\footnote{We include under instrumental noise any non-confusion sources of noise, including the detectors and photon statistics, but excluding any foreground emissions.} (\inst).
We refer to these maps as `noiseless'.

Following \citetalias{Bethermin24}, the confusion noise (\conf) is defined via an iterative 5$\sigma$ clipping process.
The mode and standard deviation of the noiseless maps are measured, positive 5$\sigma$ outliers are masked, and the standard deviation is recomputed.
This process is repeated until the standard deviation converges, defining \conf.
The classical confusion limit is then defined as 5\conf.

We model observations representative of a deep PRIMAger survey ($\sim$1500~hr~deg$^{-2}$), predicted to reach point source sensitivities (5\inst) of 124, 233, 145, 209, 294, 375~\ujy\ in PHI1, PHI2, PPI1, PPI2, PPI3, PPI4 respectively.
White Gaussian instrumental noise is added on a per pixel level, based on the predicted sensitivities.
Whilst spatially uncorrelated noise is unrealistic, we confirm our pixel noise yields the correct point source performance.
A more realistic instrumental noise model, alongside a more sophisticated map making algorithm is the subject of current and future work.

\begin{figure*}
    \centering
    \includegraphics[width=\linewidth]{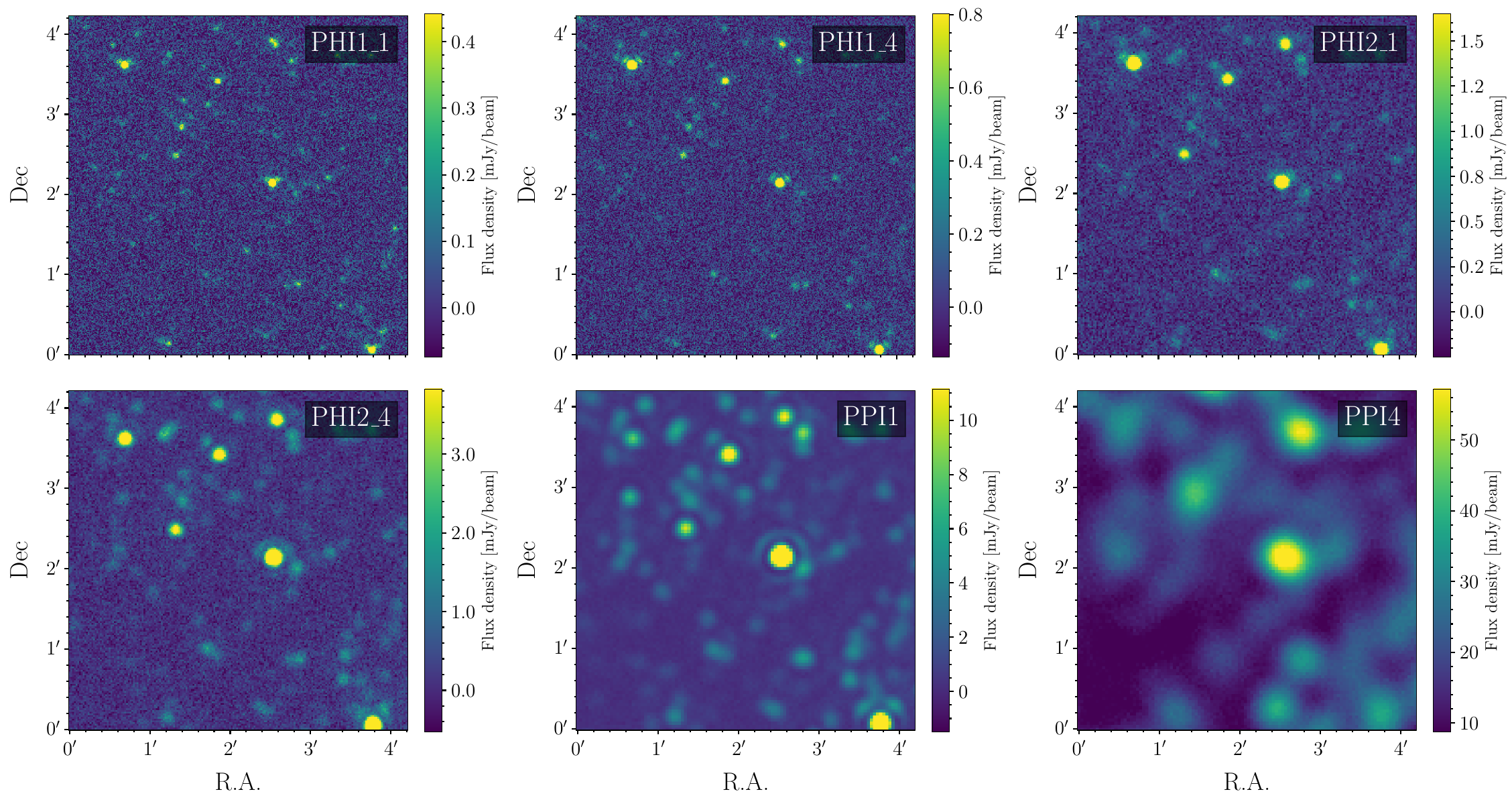}
    \caption{Cutouts of simulated observations including both instrumental and confusion noise for six representative PRIMAger channels. The transition between instrumental and confusion noise dominating is apparent. Each cutout covers an area of 4.24~$\times$~4.24$^\prime$.
    The instrumental noise is added on a per pixel level representative of a deep ($\sim$1500~hr~deg$^{-2}$) survey, as discussed in~\ref{sec:observations_mapmaking}.}
    \label{fig:noisy_cutouts}
\end{figure*}

Maps containing both confusion and instrumental noise are referred to as `noisy' hereafter, and we emphasise that they do not include any cirrus emission.
Cutouts of these maps are shown for six representative PRIMAger channels in Figure~\ref{fig:noisy_cutouts} and, unless otherwise stated, all analysis in this paper is performed on them.

In addition to these maps, which will form the basis of our main results, we produce several non-nominal maps to investigate the impact of several systematics on our flux recovery.

\subsubsection{Cirrus-contaminated maps}
\label{sec:cirrus_contaminated_maps}
\begin{figure}
    \centering
    \includegraphics[width=\linewidth]{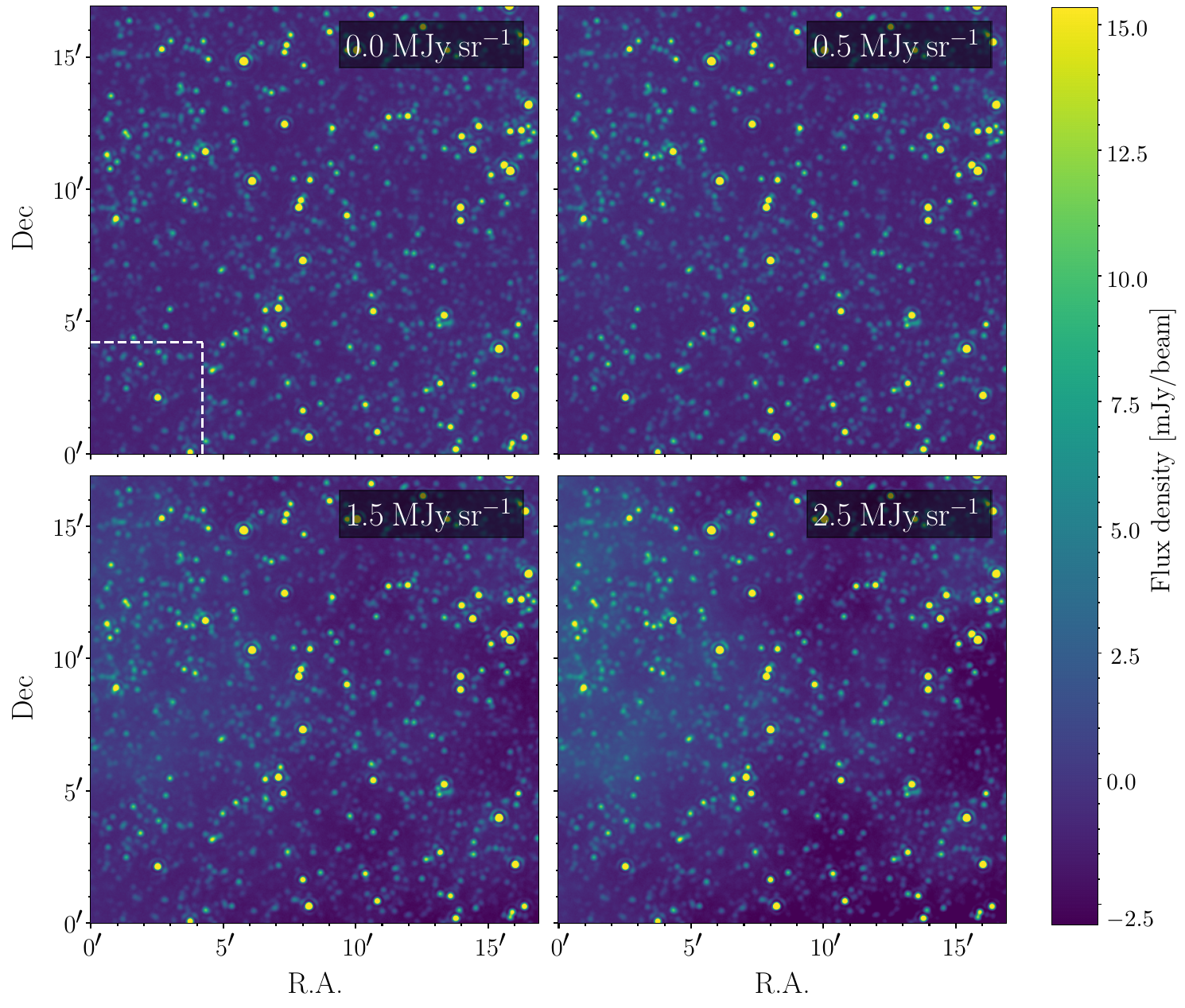}
    \caption{
    Cutouts of simulated observations in PPI1 in the presence of various cirrus levels. All cutouts share the same colour scale, are mean-subtracted, and span 17~$\times$~17$^\prime$. 
    The bounded area in the top left panel corresponds to the area shown in Figure~\ref{fig:noisy_cutouts}.
    }
    \label{fig:cirrus_cutouts}
\end{figure}

We create observations where cirrus is present at the aforementioned intensities.
We add cirrus to the noisy maps for channels redwards of PHI2\_3 (see Figure~\ref{fig:cirrus_cutouts}).

For these maps we apply a simple local background subtraction filter, modelled after
Nebuliser\footnote{\href{http://casu.ast.cam.ac.uk/surveys-projects/software-release/background-filtering}{http://casu.ast.cam.ac.uk/surveys-projects/software-release}} following the procedure outlined in \cite{HATLAS}.
Further discussion on this procedure is provided in Appendix~\ref{sec:appx_cirrus}, and the impact of cirrus on our flux recovery is discussed in Section~\ref{sec:variations}.
\\
\subsubsection{Non-nominal beams}
\label{sec:non_nominal_beams}
We consider various scenarios which would result in non-nominal telescope beams. 

Firstly, we consider the effect of wider beams on our flux recovery.
This could, for example, be the result of a smaller telescope aperture or optical aberrations.
We parametrise this following:

\begin{equation}
    \text{FWHM}_\text{degraded} = 1.1\sqrt{(\text{FWHM}_\text{nominal}^2 + c^2)},
    \label{eq:level2_beams}
\end{equation}
where $c = 3\farcs65$, and $\text{FWHM}_\text{nominal}$ and $\text{FWHM}_\text{degraded}$ is the FWHM of the nominal and degraded beams respectively.
We create these beams by broadening the nominal beams with a Gaussian to reach the target width, and refer to them as `degraded' hereafter.

Secondly, we consider that reaching the target sensitivities for a survey such as the one considered here will require stacking hundreds of independent sky scans.
Small astrometric errors in each scan, predicted to have an RMS of 0\farcs7 per axis, would result in the effective beam being broadened.
To mimic this effect, we convolve our beams with a 2D Gaussian of $\sigma$ = 0\farcs7 (FWHM = 1\farcs65), we refer to these beams as `jittered'.

Finally, we construct simplified beams to investigate the effect of miscalibrations on our flux recovery.
These beams are modelled as perfect Airy discs,

\begin{equation}
    I(r) = I_0\bigg(\frac{2J_1(kr)}{kr}\bigg)^2,
    \label{eq:airy_beam}
\end{equation}
where
$I(r)$ is the radial beam profile,
$I_0$ is the peak intensity,
$J_1$ is the first order Bessel function of the first kind,
$k$ is a band-dependent parameter controlling the beam width,
and $r$ is the radial distance from the beam centre, $r^2 = x^2 + y^2$.
For each band, $k$ is chosen such that the resulting beam has the same FWHM as the nominal beam.
Although these beams lack the realistic optical features present in our nominal beams, they allow us to independently vary the beam parameters with high fidelity (see Section~\ref{sec:xid_beam_modelling}).
These beams will be referred to as `simplified' hereafter.
A comparison of the radial profiles between the nominal and simplified beams is shown in Figure~\ref{fig:beams_2d}.

For each of the three non-nominal beam scenarios, we construct new noisy maps to run \xid.
\section{XID+}
\label{sec:xid}

\subsection{XID+: A Probabilistic De-Blender}
\label{sec:xid_general}
\xid\footnote{\url{https://github.com/H-E-L-P/XID_plus}} is a prior-based source photometry tool developed by \cite{Hurley}.
It estimates the fluxes of a collection of sources with known positions simultaneously.
The \xid model assumes that the input data ($\boldsymbol{d}$) are maps with $n_{1} \times n_{2} = M$ pixels, where the maps are formed from $N$ known sources, with flux densities $\boldsymbol{S_{i}}$ and a single global background term, $B$, accounting for unknown sources.
The point response function (PRF, $\boldsymbol{P}$) quantifies the contribution each source makes to the pixels in the map and is estimated from the point spread function (PSF) of the observatory and instrument optical system, which is assumed to be known a priori.
In this case the PRF is the same for each source and centred on the centre of the pixel containing the source, in the same way that the simulations were constructed.
The map can therefore be described as follows:

\begin{equation}
\boldsymbol{d} = \sum_{i=1}^{N} \boldsymbol{P}\boldsymbol{S_{i}} + N(0, \Sigma_{\text{inst}}) + N(B, \Sigma_{\text{conf}}),
\label{eq:xid_eq}
\end{equation}
where the two independent noise terms represent the instrumental noise and the residual confusion noise which is modelled as Gaussian fluctuations about the global background.
\xid undertakes an adaptive form of Hamiltonian Monte Carlo sampling, the no-u-turn sampler (NUTS) sampling, from this probabilistic model to obtain the full posterior. 
\cite{Hurley} utilised the Bayesian inference tool, \texttt{Stan}, however, here we build the \texttt{NumPyro} \citep{Numpyro2, Numpyro1} backend into \xid to improve speed.
It should be noted that we consider every source as a point source  within both the simulation and the modelling.
However, in real data, where extended sources are still present even in confusion dominated maps such as those from \textit{Herschel} \citep[e.g.][]{desphot2}, it would be possible to identify and flag any extended emission from the residual between the modelled map and the data map, as is further discussed in Appendix~\ref{sec:appx_failures}.
It would also be possible to provide an updated shape model instead of the PSF for any identified extended source, as each source has its own pointing matrix with \xid.

\subsection{Positional priors}
As input, \xid requires the positions of sources at which the deblending should be performed and their fluxes estimated.
These positional priors can be obtained by performing source detection on the PRIMAger maps themselves or from ancillary data such as source catalogues constructed from optical surveys at higher angular resolution.
In the following, we consider both of these methods.

\subsubsection{Euclid prior catalogues}
\label{sec:euclid_cats}
By the time PRIMA is scheduled to launch, \euclid \citep{Euclid_Overview} is expected to have completed observations of both its Wide and Deep Field surveys (with preliminary data already released: \citealt{EuclidQ1_overview}).
As such, a wealth of deep optical and NIR photometry will be available for galaxies within most regions of sky likely to be covered by PRIMA, along with robust estimates of their redshifts and stellar masses \citep{EuclidQ1_redshifts}.
Considering this, we investigate the performance of \xid for sources within the SIDES simulation which are representative of those that will be detected by the \euclid extragalactic surveys.

\begin{figure}
    \centering
    \includegraphics[width=\linewidth]{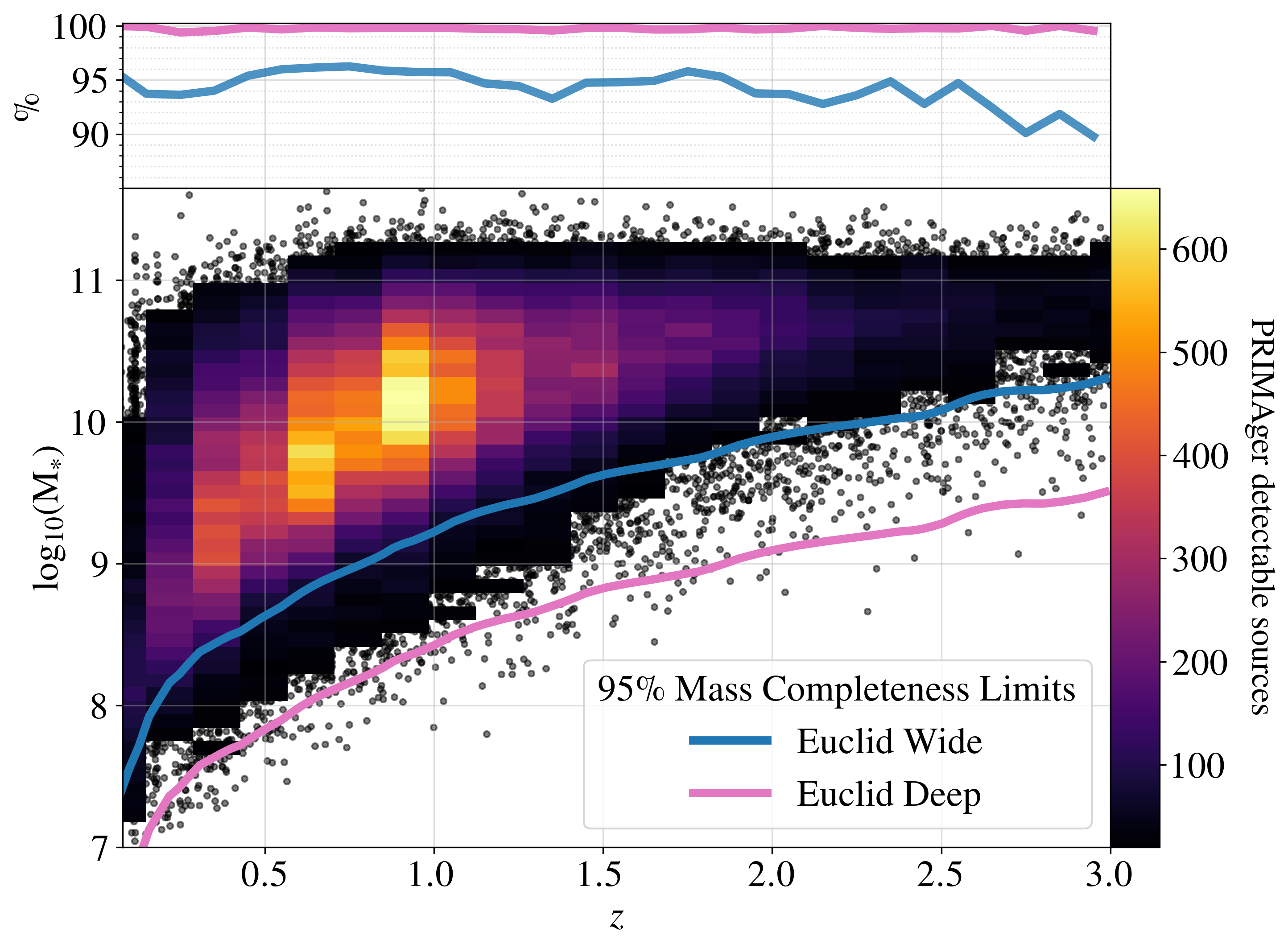}
    \caption{
    Distribution of SIDES sources which are above the point source sensitivity in at least five of the PRIMAger bands.
    \textbf{Bottom:} Stellar mass-redshift distribution of the PRIMAger detectable sources within SIDES relative to the \euclid Q1 95\% stellar mass completeness limit from \cite{EuclidQ1} (solid blue curve), representing a \euclid Wide Field survey depth and a 0.8 dex deeper \euclid Deep Field Survey (solid pink curve).
    \textbf{Top:} The percentage of PRIMAger detectable sources in SIDES which are above the mass completeness limits for the respective \euclid survey depths.
    }
    \label{fig:euclid_mass_cuts}
\end{figure}

Ideally, we would utilise simulated \euclid photometry within SIDES to determine which sources should be kept to construct a proxy, \euclid-like, source catalogue.
However, SED information for individual sources within SIDES only extends down to MIR wavelengths.
Instead, we apply a stellar mass and redshift cut to the master SIDES catalogue based on the 95\% stellar mass completeness limit derived from the \euclid Q1 data release \citep{EuclidQ1}.
This is shown in Figure~\ref{fig:euclid_mass_cuts} as the solid blue curve and is referred to as Euclid Wide since the Q1 data release is of the same depth in magnitude as the Wide Field survey depth.
It is expected that the \euclid Deep Field Survey will be $\sim$2 magnitudes deeper in apparent magnitude than the Wide survey \citep{Euclid_SR,Euclid_wide2022}. This should correspond to the stellar mass completeness being 0.8 dex deeper in stellar mass and we show this with the solid pink curve (denoted as Euclid Deep).
All SIDES sources above these respective limits form the prior source catalogues (hereafter referred to as Euclid Wide and Euclid Deep) for which the deblending will be performed.
The stellar mass completeness limit of \cite{EuclidQ1} was only derived up to $z=3$. Subsequently, we have extrapolated this relation for higher redshifts by fitting a broken power law. However, there are very few sources at these high redshifts within SIDES which significantly contribute flux to the PRIMAger maps. Therefore, any error in this extrapolation will have a negligible effect.

The exact positions of the sources within the simulated maps are used. However, we explore the impact of positional offsets on the modelling accuracy in Section~\ref{sec:variations}.

The bottom panel of Figure~\ref{fig:euclid_mass_cuts} also shows the stellar mass vs redshift distribution of the SIDES sources within the 2~\sqdeg light-cone which are above the point source sensitivity of the considered PRIMAger survey depth in at least five of the PRIMAger bands.
These sources are effectively the main contributors of flux to the PRIMAger maps and are thus referred to as PRIMAger-detectable sources.
As the top panel of Figure~\ref{fig:euclid_mass_cuts} shows, 90--95\% of these sources are above the Euclid Wide stellar mass limit, and therefore are present in the Euclid Wide prior source catalogue, out to $z \sim 3$.
For the Euclid Deep prior catalogue, effectively all PRIMAger-detectable sources are present. 

\begin{figure}
    \centering
    \includegraphics[width=\linewidth]{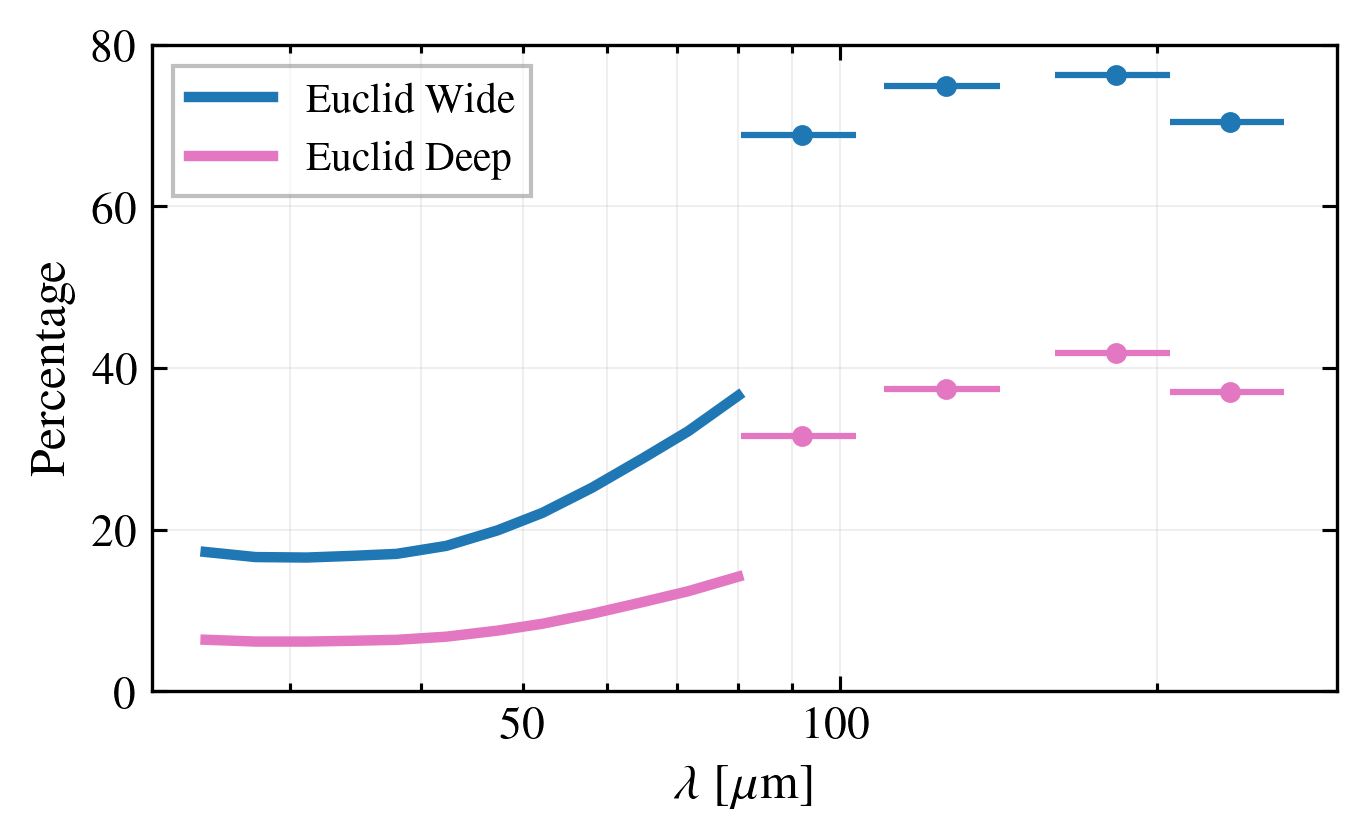}
    \caption{Percentage of sources within the Euclid prior catalogues which are above the instrument point source sensitivity at each of the PRIMAger channels for the considered survey. At the shortest wavelengths, less than 20\% of the sources within the prior catalogues are detectable above the point source sensitivities and significantly contributing flux to the maps.}
    \label{fig:euclid_cats_percentage_wvs}
\end{figure}

Due to the impressive depths in stellar mass reachable with \euclid, there can be significant fractions of sources within the prior catalogues which are not significantly contributing flux in a given PRIMAger channel (e.g. quiescent galaxies).
Figure~\ref{fig:euclid_cats_percentage_wvs} shows the percentage of sources which are above the point source sensitivity of the considered PRIMAger survey for a given channel within the two \euclid-based prior catalogues.
At the shortest wavelengths, less than 20\% of sources present in the prior catalogues are actually significantly detectable (i.e. above the point source sensitivity) in the simulated maps themselves.
The remaining sources will effectively be adding more complexity and degeneracy to the \xid modelling.
In order to mitigate this, as concluded by \citetalias{Donnellan24}, some form of prior knowledge on the fluxes of the sources is required and is discussed in Section~\ref{sec:flux_priors}.

\subsubsection{Blind source detection}
\label{sec:blind_pos_prior}
PRIMA is scheduled to launch at a time where a plethora of ancillary data will be available, such as from \euclid (as described above), the \textit{Nancy Grace Roman Space Telescope} \citep{Roman_Overview}, the Vera Rubin Observatory \citep{Rubin_Overview}, or radio surveys, such as from the Square Kilometre Array \citep[SKA;][]{SKA_Overview}.
Despite this, here we aim to demonstrate that PRIMAger is capable of achieving its scientific goals in the absence of this data, in a fully self-contained way; detecting and analysing sources entirely from PRIMAger observations.
For this purpose, we perform a blind source detection.

In observations where source confusion is a significant contributor to the total map noise (\total; $\text{\total}^2=\text{\inst}^2+\text{\conf}^2$), 
convolving with the PSF no longer optimises the signal-to-noise of isolated point sources.
Instead, a matched (or Wiener) filter, created taking into account the noise properties of the map, is optimal \citep{Chapin11}.
The blind detection process is outlined in \citetalias{Donnellan24}, though a short summary is included here. 

For each channel, a matched filter is created following the \citet{Chapin11} prescription.
Source detection is run on filtered maps using the \texttt{find\_peaks} method provided by \texttt{Photutils} \citep{Photutils3.0.0}.
A source is determined to be present in a pixel if its value is the local maximum within a 5~$\times$~5~pixel region centred on it.
Since the maps are in Jy~beam$^{-1}$, source fluxes can be estimated by directly reading their pixel value.

For confusion dominated channels, the application of the matched filter introduces distinct ringing features at known distances and intensities.
These rings are likely to be detected as sources by the peak-finder algorithm if originating from sufficiently bright sources.
To remove these false positives, we examine all sources which are expected to produce ringing features above \total.
Peaks found in the expected ringing region, whose flux is less than five times the predicted ringing, are considered contaminated and are thus removed.

Similarly to \citetalias{Donnellan24}, we cut the catalogue based on an observed flux threshold which corresponds to a 95\% purity catalogue, where purity is defined as the fraction of detected peaks which have a counterpart in the true catalogue.
The true counterpart is the brightest source which satisfies the following criteria:
(i) the positional offset between the detected peak and the true source, $d_\text{off}$, satisfies $d_\text{off} \leq$ \text{FWHM}/2,
and (ii) the ratio of the observed and true flux, S$_{\text{obs}}$/S$_{\text{true}}$, satisfies 0.5 $\leq$ S$_{\text{obs}}$/S$_{\text{true}}\leq$ 2.

This process is repeated for each channel, and the individual catalogues are merged to give a list of unique positional priors.

\subsection{Flux priors}
\label{sec:flux_priors}
Within the \xid framework, it is possible to utilise prior flux knowledge of specific sources to augment the performance of the modelling.
This can be done by either using the prior flux knowledge to inform which sources will be kept in the prior source catalogue or by placing constraints on the source fluxes within the modelling itself. 

The former was explored by \cite{Wang24} whereby a neural network-based CIGALE \citep{CIGALE} emulator was used to predict \herschel/PACS and SPIRE fluxes from \spitzer/MIPS 24~\um photometry.
These predicted fluxes then informed which sources were to be kept and passed to \xid. The actual modelling was then performed with uniform prior distributions on the fluxes for the remaining sources. 

Another approach is to retain every source in the prior list and instead inform the prior distribution on the source fluxes within the modelling of \xid.
\citetalias{Donnellan24} investigated this method by placing a Gaussian prior distribution on simulated source fluxes which were centred on the true flux of the given source and a standard deviation also equal to the true flux.
This was designed to mimic a procedure which would utilise predicted FIR fluxes from SED-fitting data obtained at higher spatial-resolutions.
\citetalias{Donnellan24} found that significant gains in flux modelling accuracy could be obtained when applying these flux priors for suitably high-density prior source catalogues, such as those based on \euclid data.

In the following we consider an extension of this method which fully utilises the hyperspectral imaging capabilities of PRIMAger to obtain prior flux information. 

\subsubsection{XID+stepwise}
\label{sec:stepwise}

For the deep survey considered here, the first hyperspectral band of PRIMAger, PHI1, will be instrumental noise limited due to the increased angular resolution at these shorter wavelengths significantly suppressing confusion noise.
Therefore source flux measurements obtained across this band can be used to inform the prior flux constraints within the \xid modelling at the confused longer wavelengths. 

We propose a methodology (hereafter referred to as \stepwise) which steps through each PRIMAger channel in ascending wavelength order (descending angular resolution) and performs the \xid deblending with prior flux distributions constructed from the measured fluxes of the previous step.
Specifically, the prior probability of the flux of source $i$, in a given channel $j$, is normally distributed around a mean $\mu$ with width $\sigma$:

\begin{equation}
p(S_{i,j}) \sim 
\mathcal{N}(\mu, \sigma).
\label{eq:stepwise_1}
\end{equation}

Both $\mu$ and $\sigma$ are set to the median posterior flux from the previous \xid fit at the next-shortest wavelength channel, $S_{i,j-1}$, multiplied by some channel-dependent correction factor, $C_{j}$: 
\begin{equation}
    \mu = \sigma = C_{j} S_{i,j-1}.
\label{eq:stepwise_2}
\end{equation}

For the very first channel at 25~\um ($j=0$), the prior on the source fluxes is uniformly distributed between zero and the brightest pixel it contributes to.
For all subsequent channels, the width of the prior is set to equal the mean in order to ensure it is wide enough to capture large changes in flux between bands due to, for example, the presence of PAH emission lines.
This distribution is similarly truncated between zero and the brightest pixel the source contributes to.

The wavelength-dependent correction factor takes into account the general change in SED between the observed wavelengths.
It is determined by the average ratio between fluxes at a given wavelength and the previous wavelength for all galaxies within the Euclid prior catalogues:

\begin{equation}
    C_{j} = \frac{1}{N}\sum^{N}\frac{S_{i,j}}{S_{i,j-1}},
\label{eq:stepwise_3}
\end{equation}
which gives:

\begin{equation}
C_{j} =
\begin{cases}
1.0; & j < 6 \ (\text{PHI1}) \\
1.3; & 6 \leq j < 12 \ (\text{PHI2}) \\
2.0; & 12 \leq j < 15 \ (\text{PPI1-PPI3}) \\
1.0; & j = 15 \ (\text{PPI4}).
\end{cases}
\label{eq:stepwise_4}
\end{equation}

On a source-by-source basis this correction factor has a large scatter and is redshift-dependent.
However, here we conservatively consider a population-based value as to minimally inform the SED variation between bands.
This ensures that the method is not overly sensitive to the SED templates within the chosen simulation.

In addition, we also consider the benefit of supplying some weak prior flux information at the first step, PHI1\_1, and then applying the \stepwise methodology as outlined above.
In these runs, we adopt the toy flux prior model from \citetalias{Donnellan24}, whereby the prior distribution on a source's flux in the PHI1\_1 band is a Gaussian with a mean and width equal to the true flux of the source in this band.
This provides weak flux information at the initial step (as opposed to none in the regular \stepwise setup) which could be obtained from, for example, applying SED-fitting of ancillary optical/NIR photometry.
We refer to this process, in which a prior is given at the first step followed by the standard stepwise methodology, as `informed stepwise' hereafter.

We perform three main runs in this work, summarised in Table~\ref{tab:main_runs}.
Our `Fiducial' run consists of sources within the Euclid Wide positional prior and the newly introduced stepwise methodology, providing no flux information in the first band.
Additionally, we run a `Deep' run, consisting of the Euclid Deep positional prior and an informed stepwise flux prior.
Finally, we perform a `Blind' run which is purely PRIMAger selected, with sources detected during blind source detection and a completely uninformed flux prior.

\begin{deluxetable*}{lccccc}
\tablecaption{Specifications of XID+ runs}

\tablehead{
\colhead{Run} &
\colhead{Positional} &
\colhead{Flux} &
\multicolumn{3}{c}{Source Density}
\\
\colhead{Name} &
\colhead{Prior} &
\colhead{Prior} &
\colhead{PHI1\_1}&
\colhead{PHI2\_1} &
\colhead{PPI1}
}

\startdata
Fiducial & Euclid Wide      & Stepwise (Uniform @ 25~\um)   & 0.11 & 0.37 & 0.91 \\
Deep     & Euclid Deep      & Stepwise (Prior @ 25~\um)     & 0.31 & 1.00 & 2.47 \\
Blind    & Blind Detection  & Uniform                       & 0.05 & 0.16 & 0.39
\enddata

\tablecomments{
A summary of the main runs included in this work.
The positional prior specifies the locations at which \xid should perform the deblending.
Euclid Wide and Euclid Deep refer to the stellar mass-redshift cut achievable by the \euclid Wide and Deep Field surveys respectively (Section~\ref{sec:euclid_cats}).
Blind refers to the source detection performed in a self-contained way (Section~\ref{sec:blind_pos_prior}).
Flux prior determines the prior flux information of individual sources given to \xid.
Uniform uses a constant distribution limited to between zero and the brightest pixel a source contributes to, whereas stepwise refers to the process of obtaining prior flux information for a given channel from the measurements made in the previous channel (Section~\ref{sec:stepwise}).
Source density refers to the number of sources present in the catalogue per beam, quoted for the PHI1\_1, PHI2\_1, and PPI1 channels.
}
\label{tab:main_runs}
\end{deluxetable*}

\subsection{Systematic errors in the Point Response Function}
\label{sec:xid_beam_modelling}
To model how each source contributes flux to nearby pixels, \xid estimates the PRF from an input PSF.
Typically, we give \xid the same PSF used to generate the maps.
We expect to accurately measure the beam to within a few percent both prior to launch and in flight.

To quantify the impact of beam miscalibrations, we run \xid on maps generated using the simplified Airy disc beams described in Section~\ref{sec:non_nominal_beams}.
The maps are kept the same for each run, varying only the model beam assumed by \xid.
We perform four separate \xid runs with beams derived from the Airy disc profile described in Eq.~\ref{eq:airy_beam}, comparing performance when providing \xid:
(i) the true beam (i.e. the same beam used to generate the maps),
(ii) a beam broadened by 5\% ($r^\prime = 1.05r$),
(iii) a beam made elliptical by broadening one axis by 5\% ($r^\prime = \sqrt{(x/1.05)^2 + y^2}$),
and (iv) a beam with 50\% stronger ringing, implemented as
\begin{equation}
    I_0^\prime =
    \begin{cases}
        I_0; & r<d_0\\
        1.5I_0; & r\geq d_0,
    \end{cases}
\end{equation}
where $d_0$ is the radius of the first Airy zero.

The results from this analysis is discussed in Section~\ref{sec:variations}.
\\
\subsection{Modelling performance metrics}
\label{sec:metrics}
In order to quantify the flux accuracy of the \xid modelling and to determine the level down to which we can reliably measure source fluxes from the PRIMAger maps, we adopt the `limiting flux' statistic of \citetalias{Donnellan24}.
It quantifies the deviation of the extracted fluxes, $S_{\text{obs}}$, from the true fluxes, $S_{\text{true}}$, within bins of true flux using the median absolute deviation, $\sigma_{\text{MAD}}$:

\begin{equation}
\sigma_{\text{MAD}}(S_{\text{true}}) = \alpha \cdot \text{Med}\left(\frac{\Delta S}{S_{\text{true}}} - \text{Med}\left(\frac{\Delta S}{S_{\text{true}}}\right)\right),
\label{eq:mad}
\end{equation}
where $\Delta S = S_{\text{obs}} - S_{\text{true}}$ and $\alpha = 1.4826$ (which scales the MAD to a Gaussian).
The limiting flux, $S_{\text{lim}}$, is then defined as the flux at which $\sigma_{\text{MAD}}$ equals 0.2:

\begin{equation}
S_{\text{lim}} = S_{\text{true}}\bigg|_{\sigma_{\text{MAD}} = 0.2}.
\label{eq:limiting_flux}
\end{equation}
This corresponds to the true flux at which the median deviation of the observed fluxes from the true values equals 20\% of the true flux.
Noise within the limiting flux statistic can arise due to the fact that it depends upon the binning of sources by their true flux values.
As such, bootstrapping will be performed to estimate any uncertainty in the limiting flux.
\section{Results}
\label{sec:results}

\subsection{Main results}
Differences in our results from those in \citetalias{Donnellan24} arise primarily from the introduction of more realistic instrument parameters (beams, target sensitivity in noise model, etc.).
For example, the sidelobe pattern has an impact in increasing the confusion noise.
Nonetheless the overall performance remains similar, with the PHI1 channel being instrumental noise dominated, whilst confusion dominates from PHI2 onwards.

For all three main runs, we run \xid on a $\sim$0.2~\sqdeg contiguous area of our simulated field.
To limit computational cost, \xid is independently applied to 256 $\sim$3~arcmin$^2$ tiles, and comprises 4 MCMC chains with 1,000 warm-up and 1,000 regular samples.
Further details on the tiling process, alongside a summary of the computational cost and performance, are provided in Appendix~\ref{sec:appx_compute}.
The output from \xid is the full posterior distribution for the flux in a given channel for each source in the prior catalogue, including the correlation between sources.
The flux of a source is quoted as the median of its marginalised posterior flux distribution.
For each tile, posteriors for a background value (B) and residual confusion noise ($\Sigma_\text{conf}$) are also obtained.
These parameters capture flux from sources that may be missing from the prior catalogue or from any foreground contamination.

\subsubsection{Fiducial run: using Euclid Wide priors}
\label{sec:euclid_wide_results}
Our Fiducial run started from the Euclid Wide prior source catalogue.
We followed the \stepwise method described in Section~\ref{sec:stepwise}, modelling the highest-angular resolution map at 25~\um without prior flux information and modelling the map of each subsequent channel with the prior flux information derived from the posterior of the previous channel, following Equations~\ref{eq:stepwise_1}-\ref{eq:stepwise_4}.

\begin{figure*}
    \centering
    \includegraphics[width=\linewidth]{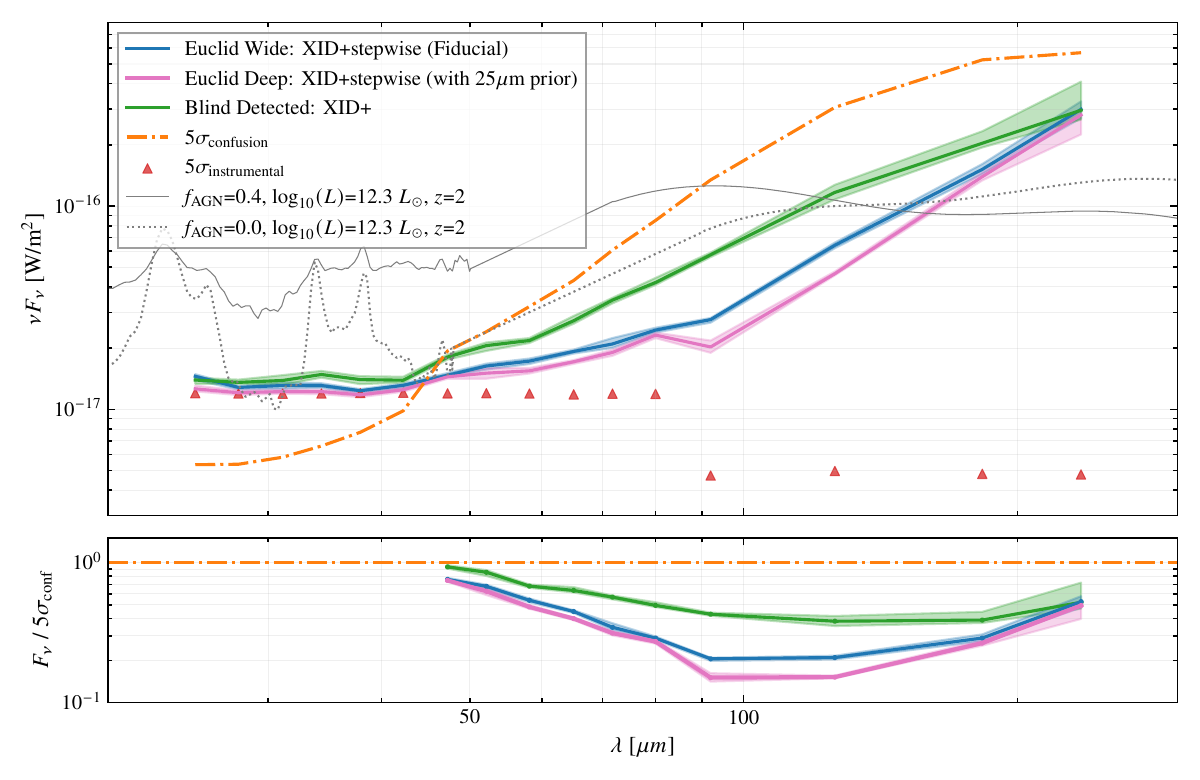}
    \caption{
    \textbf{Top:} Limiting flux density as a function of wavelength from 25--235~\um for \xid deblending of positional prior catalogues employing expected sensitivities for the \euclid Wide or Deep surveys.
    The instrumental sensitivities illustrated (red triangles) correspond to the anticipated sensitivities of the PRIMAger Deep Survey.
    The confusion curve (dash-dot orange) illustrates the confusion noise computed for our expected beam and the SIDES sky model.
    The blue and magenta lines show the expected performance from our \stepwise model using Euclid prior catalogues with the anticipated sensitivities for the \euclid Wide and Deep fields respectively, with the shaded regions showing the 1$\sigma$ uncertainty in this determination.
    For the Euclid Deep catalogue, a weak flux prior at 25~\um is supplied to reach the point source sensitivity.
    The limiting flux from \xid with no prior flux information for blind detected positional priors is also shown (green).
    The solid and dotted black curves show reference SEDs \citep{Kirkpatrick2015} for two galaxies at $z=2$ with L$^{*}$ IR luminosity \citep{Magnelli2013} and AGN fractions of 0.4 and 0 respectively.
    The intersection between these curves and the blue and magenta lines show out to what PPI channel we expect to be able to recover fluxes.
    This shows that $z=2$ L$^{*}$ galaxies will be recovered out to PPI2.
    \textbf{Bottom:} Limiting flux density reached by each run relative to the 5$\sigma$ confusion limits for the confusion-dominated channels.
    }
    \label{fig:euclid_wide_limiting_flux}
\end{figure*}

Using Equations~\ref{eq:mad} and~\ref{eq:limiting_flux}, the limiting flux threshold, corresponding to a measured flux accuracy of 20\% (5$\sigma$), is determined for each of the PRIMAger bands in the Fiducial run and are shown in Figure~\ref{fig:euclid_wide_limiting_flux} and tabulated in Table~\ref{tab:results_stats}. These are compared to the classical confusion limits, as calculated in Section~\ref{sec:observations_mapmaking}, and the point source sensitives of the simulated deep survey. 

For all bands in PHI1 ($\lambda <$ 45~\um), the instrumental noise dominates confusion and \xid is able to recover accurate fluxes down to within 3--21\% above the point source instrumental sensitivities.
The largest discrepancy is at PHI1\_1.
Even after accounting for noise via bootstrapping (shown by the blue shaded region in Figure~\ref{fig:euclid_wide_limiting_flux}), the limiting flux at PHI1\_1 is still slightly above the point source sensitivity.
This is due to the large number of sources in the prior catalogue.
Many of these do not contribute significant flux to the map but do add degeneracy to the modelling (as shown in Figure~\ref{fig:euclid_cats_percentage_wvs}), with flux from brighter sources misattributed to fainter sources.
With only positional information and no prior flux information in this band to break the degeneracies, this results in a slightly elevated noise relative to the instrumental noise in the map.
We return to this issue in Section~\ref{sec:flux_knowledge_impact}.

For all channels in PHI2 and PPI ($\lambda \geq$ 45~\um), confusion noise dominates the instrumental noise.
The limiting flux of our Fiducial run reaches below the classical confusion limit for all of these confusion-dominated channels.
For PHI2, the limiting flux is a factor of 1.3--3.5 below the classical confusion limit between 45--84~\um.
The biggest gains are achieved for PPI, where the limiting flux is factors of 4.8, 4.8, 3.5, and 1.9 below the confusion limits for PPI1 through PPI4 (92--235~\um) respectively.

\begin{deluxetable*}{lcccc}
\tablecaption{XID+ Limiting Flux}

\tablehead{
\colhead{\begin{tabular}{c}{Channel}\end{tabular}} &
\colhead{\begin{tabular}{c}$\lambda_{\text{peak}}$\end{tabular}} &
\colhead{\begin{tabular}{c}Fiducial\\Limiting Flux\end{tabular}} &
\colhead{\begin{tabular}{c}Deep\\Limiting Flux\end{tabular}} &
\colhead{\begin{tabular}{c}Blind\\Limiting Flux\end{tabular}}\\
\colhead{} &
\colhead{{[\um]}} &
\colhead{{[\ujy]}} &
\colhead{{[\ujy]}} &
\colhead{{[\ujy]}}
}

\startdata
PHI1\_1 & 25.0 & \valasym{121}{3}{5}            & \valasym{105}{4}{3}               & \valasym{116}{3}{5} \\
PHI1\_2 & 27.8 & \valasym{119}{2}{3}            & \valasym{112}{3}{3}               & \valasym{126}{4}{2} \\
PHI1\_3 & 31.1 & \valasym{136}{5}{5}            & \valasym{127}{3}{3}               & \valasym{144}{9}{7} \\
PHI1\_4 & 34.3 & \valasym{150}{4}{4}            & \valasym{140}{3}{4}               & \valasym{170}{7}{6} \\
PHI1\_5 & 37.9 & \valasym{156}{3}{4}            & \valasym{149}{5}{4}               & \valasym{177}{7}{9} \\
PHI1\_6 & 42.3 & \valasym{185}{4}{6}            & \valasym{176}{7}{4}               & \valasym{196}{8}{5} \\
PHI2\_1 & 47.3 & \valasym{232}{6}{3}            & \valasym{228}{4}{5}               & \valasym{285}{8}{8} \\
PHI2\_2 & 52.1 & \valasym{284}{8}{11}           & \valasym{262}{15}{16}             & \valasym{358}{13}{21} \\
PHI2\_3 & 58.2 & \valasym{335}{10}{11}          & \valasym{300}{10}{9}              & \valasym{423}{12}{12} \\
PHI2\_4 & 65.0 & \valasym{418}{9}{10}           & \valasym{372}{7}{9}               & \valasym{592}{33}{20} \\
PHI2\_5 & 71.8 & \valasym{502}{35}{22}          & \valasym{456}{15}{17}             & \valasym{825}{22}{30} \\
PHI2\_6 & 80.0 & \valasym{655}{18}{18}          & \valasym{619}{18}{21}             & \valasym{1,120}{60}{30} \\
PPI1    & 92.0 & \valasym{848}{16}{29}          & \valasym{622}{47}{40}             & \valasym{1,770}{50}{50} \\
PPI2    & 126  & \valasym{2,700}{80}{90}        & \valasym{1,950}{50}{40}           & \valasym{4,910}{440}{380} \\
PPI3    & 183  & \valasym{9,250}{580}{480}      & \valasym{8,520}{390}{360}         & \valasym{12,400}{1,800}{600} \\
PPI4    & 235  & \valasym{23,400}{2,200}{2,200} & \valasym{22,000}{1,100}{4,400}    & \valasym{23,200}{8,900}{2,300} 
\enddata

\tablecomments{
Limiting flux for each \xid run in the 16 PRIMAger channels, as presented in Figure~\ref{fig:euclid_wide_limiting_flux}.
The channel name and central wavelength are given, with additional specifications for each channel presented in Table~\ref{tab:filter_stats}.
The three runs are summarised in Table~\ref{tab:main_runs}, and are: the Euclid Wide catalogue with \stepwise (Fiducial), the Euclid Deep catalogue with informed \stepwise (Deep), and the blind detected catalogue with \xid (Blind).
For each of the runs, the limiting flux is determined in each channel following Equations~\ref{eq:mad}--\ref{eq:limiting_flux}.
Quoted values are the median of 1,000 bootstrapped samples, with uncertainties given by the 16$^\text{th}$ and 84$^\text{th}$ percentiles, and are given to three significant figures.
}
\label{tab:results_stats}
\end{deluxetable*}

\begin{figure}
    \centering
    \includegraphics[width=\linewidth]{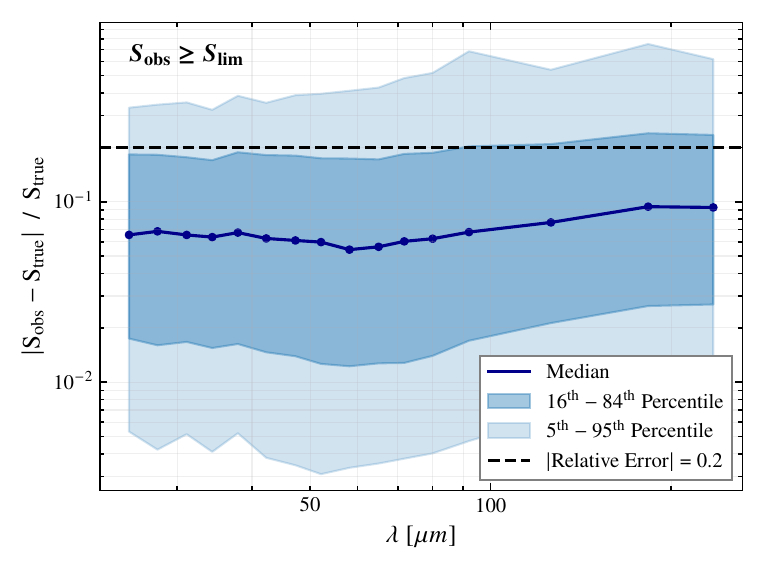}
    \caption{
    Distribution of the absolute relative errors between the observed and true fluxes from \stepwise for all sources with an observed flux above the limiting flux at the given wavelength for the Euclid Wide prior source catalogue.
    For all PRIMAger channels, the median absolute relative error is well below the threshold of 0.2 (dashed black line) which represents an observed flux accuracy of 20\%.
    The 84$^{\text{th}}$ percentile is also below this threshold for all PHI channels.
    }
    \label{fig:euclid_wide_rel_errs}
\end{figure}

In order to validate the robustness of the limiting flux statistic, we examine the distribution of the absolute relative errors between the observed and true fluxes for sources with an observed flux above the limiting flux.
These are shown for all channels for our Fiducial run in Figure~\ref{fig:euclid_wide_rel_errs}.
As can be seen, the median absolute relative error is well below the threshold of 0.2, representing an observed flux accuracy of 20\%, for all channels.
In fact, the median error is $\sim6\%$.
Moreover, the 84$^\text{th}$ percentile is also below 20\% for all PHI channels.

There is however a bias of underestimating fluxes for faint sources and longer wavelength.
This bias can be calibrated and thus corrected and is discussed in Appendix~\ref{sec:appx_xid_biases}.
This underestimation is $\lesssim5\%$ at the limiting flux for PHI1, increasing up to $\sim$15\% for PPI.
The uncertainty in the measured \xid fluxes, derived from the spread in their posterior distributions from \xid, are also discussed in Appendix~\ref{sec:appx_xid_biases}. 
A selection of extracted SEDs are compared to the true SEDs for sources with a range of redshifts and LIRs in Appendix~\ref{sec:appx_xid_seds}, as well as a discussion on the rates of pathological failures in the source flux modelling.

\subsubsection{Using Euclid Deep priors}
\label{sec:euclid_deep_results}
Figure~\ref{fig:euclid_wide_limiting_flux} also shows the limiting flux reached when applying \stepwise to the Euclid Deep prior source catalogue along with some weak prior flux information at the first step, PHI1\_1, as outlined at the end of Section~\ref{sec:stepwise}.
The high density of sources in the Euclid Deep catalogue (which contains $\sim$2.8 times more sources than Euclid Wide) necessitates the use of prior flux information in the first step of \stepwise at PHI1\_1 for the limiting flux to reach the point source sensitivity.
This has a knock-on effect on the limiting flux depths in subsequent bands due to the \stepwise methodology.
The very modest amount of prior flux information introduced at 25~$\mu$m leads to greater sensitivity in all wavelengths compared to the Fiducial run, resulting in up to a factor of $\sim$1.4 deeper limiting flux at PPI1, despite the larger number of catalogue sources.
The Euclid Deep catalogue naturally has a significantly lower fraction of sources which contribute significant flux to the maps across all the bands (as shown by Figure~\ref{fig:euclid_cats_percentage_wvs}).
Hence, without the additional flux information at the initial step, these extra sources would add degeneracies to the modelling and cause the limiting flux to lie above the point source sensitivity, an effect similar to that seen at PHI1\_1 in our Fiducial run.
We further explore the impact of including and excluding prior flux knowledge for PHI1\_1 in Section~\ref{sec:flux_knowledge_impact}.

\subsubsection{Using Blind priors}
For our Blind run, we restrict ourselves to data obtainable by PRIMAger, not utilising any ancillary data neither for positional nor flux priors.
The positional prior arise from objects detected through matched filtering, and the flux prior is modelled in \xid with a uniform distribution.
Even in this scenario, \xid is capable of reaching close to the instrumental noise for PHI1, averaging 17\% above the noise floor.
Additionally, the limiting flux is up to a factor of 2 and 2.6 deeper than the confusion limit for PHI2 and PPI respectively.

\subsection{Improvement provided by \stepwise}
\label{sec:flux_knowledge_impact}

In order to investigate the impact on recovered flux accuracy when applying prior flux knowledge via \stepwise, we also run \xid without any flux information at any of the bands (i.e. a uniform flux prior) for both Euclid positional prior catalogues. 

\begin{figure}
    \centering
    \includegraphics[width=\linewidth]{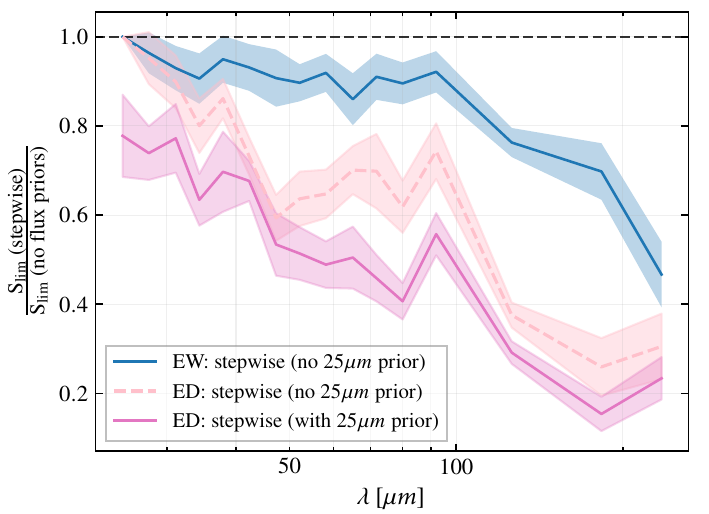}
    \caption{
    The ratios of the limiting flux reached by \xid when including (via the stepwise methodology) and excluding prior flux knowledge (where only a uniform flux prior distribution is used) for both Euclid catalogues.
    The ratios of the limiting fluxes of these runs are shown by the blue solid curve and pink dashed curve for the Euclid Wide and Deep catalogues respectively.
    Significant gains in the limiting flux are achieved when applying \stepwise compared to the uniform run for both catalogues by up to a factor of four at the longest wavelengths.
    For the Euclid Deep catalogue, we also run \stepwise with a weak prior 25~$\mu$m flux for the PHI1\_1 band.
    The ratio of the limiting flux from this run and that of the uniform run for the Euclid Deep catalogue is shown by the solid magenta curve.
    Comparing the solid magenta to the dashed pink curve highlights the additional gains at all wavelengths from supplying weak prior flux information at 25~$\mu$m for the Euclid Deep catalogue.
    }
    \label{fig:flux_knowledge_impact}
\end{figure}

For the Euclid Wide catalogue, we compare the limiting flux reached by \stepwise (i.e. our Fiducial run) and that reached by the run with no prior flux information (referred to as the `uniform' run) and show the ratio between these two as the solid blue curve in Figure~\ref{fig:flux_knowledge_impact}.
At PHI1\_1, this ratio is unity due to the first step in \stepwise assuming no prior flux information and applies the same uninformative, uniform distribution.
For all subsequent bands, however, the limiting flux is improved by incorporating the flux information acquired from the previous band through \stepwise.
For channels PHI1\_2--PPI1, the limiting flux is improved by $\sim$5--15\%, with even further improvements of $\sim$20--55\% for the most confused channels (PPI2--PPI4).

In all cases, but particularly for the high source density Euclid Deep catalogue, leveraging flux information acquired from the previous band through the stepwise methodology leads to substantial gains in the limiting flux.
This is shown by the pink dashed curve in Figure~\ref{fig:flux_knowledge_impact}, where the limiting flux reached in the confusion-dominated PHI2 channels (45--84~\um) using \stepwise are $\sim$20--40\% below that of running \xid.
For the two longest (and most confused) wavelength bands, PPI3 and PPI4, this leads to very large gains, decreasing the limiting flux by $\sim$70\% relative to that of \xid.

It is worth noting the sharp peak seen at PPI1. This is due to a combination of the significantly better point source sensitivity in PPI (Figure~\ref{fig:euclid_wide_limiting_flux}), and a large increase in the number of sources within the prior source catalogue which significantly contribute flux to the map in this band (see Figure~\ref{fig:euclid_cats_percentage_wvs}).
This leads to good performance from \xid in this channel and therefore less value is added from applying \stepwise.

We also investigate the impact of weak prior flux information at the first step of \stepwise, PHI1\_1, as opposed to using an uninformative flat prior.
This is motivated by the behaviour of the limiting flux relative to the point source sensitivity at PHI1\_1 in our Fiducial run as discussed in Section~\ref{sec:euclid_wide_results}.
Working with the Euclid Deep catalogue, we obtain a $\sim$20\% improvement in the limiting flux depth at PHI1\_1 with an informed \stepwise.
This is illustrated by the first data point of the solid magenta curve in Figure~\ref{fig:flux_knowledge_impact}.
Application of this initial prior also consistently improves the limiting flux reached in the rest of the bands. 

For the Euclid Wide catalogue, however, supplying weak prior flux information at PHI1\_1 has a negligible effect across all channels, except for PHI1\_1 itself.
This is due to the fact that the limiting flux is already very close to the point source sensitivities of PHI1 in the Fiducial run, therefore the flux modelling is already performing as well as possible and is ultimately limited by instrumental sensitivity.

\subsection{Variations from nominal}
\label{sec:variations}

\begin{deluxetable}{lccc}
\tablecaption{Specifications of varied runs}

\tablehead{
\colhead{Variation Name} &
\colhead{\Ihundred [\mjysr]} &
\colhead{Map Beam} &
\colhead{XID+ Beam}
}

\startdata
Low Cirrus      & 0.5 & Nominal     & Nominal   \\
Medium Cirrus   & 1.5 & Nominal     & Nominal   \\
High Cirrus     & 2.5 & Nominal     & Nominal   \vspace{5pt}\\

Degraded        & 0.0 & Degraded    & Degraded  \\
Jittering       & 0.0 & Jittered    & Jittered  \vspace{5pt}\\

Broader         & 0.0 & Simplified  & Broader   \\
Elliptical      & 0.0 & Simplified  & Elliptical\\
Ringing         & 0.0 & Simplified  & Ringing
\enddata

\tablecomments{
Summary of the different runs done to investigate the effects of systematics on our flux recovery.
We consider low, medium, and high levels of cirrus contamination, corresponding to cirrus intensities (\Ihundred) of 0.5, 1.5, and 2.5~\mjysr respectively (Sections~\ref{sec:modelling_cirrus} and~\ref{sec:cirrus_contaminated_maps}).
We consider two scenarios, `degraded' and `jittered', where the telescope beam is broader than currently assumed (Section~\ref{sec:non_nominal_beams}).
Finally, we perform three runs to see the effect of providing \xid with a beam that differs from the one used to create the maps, individually checking for the effect of beam width, ellipticity, and ringing strength (Section~\ref{sec:xid_beam_modelling}).
All runs were performed with the Euclid Wide positional prior, and a uniform flux prior.}
\label{tab:variation_runs}
\end{deluxetable}

We investigate three systematic effects which may affect PRIMAger observations:
(i) the presence of cirrus,
(ii) a telescope beam larger than currently expected,
and (iii) the impact of miscalibration in the beam on \xid modelling.

\begin{figure}
    \centering
    \includegraphics[width=\linewidth]{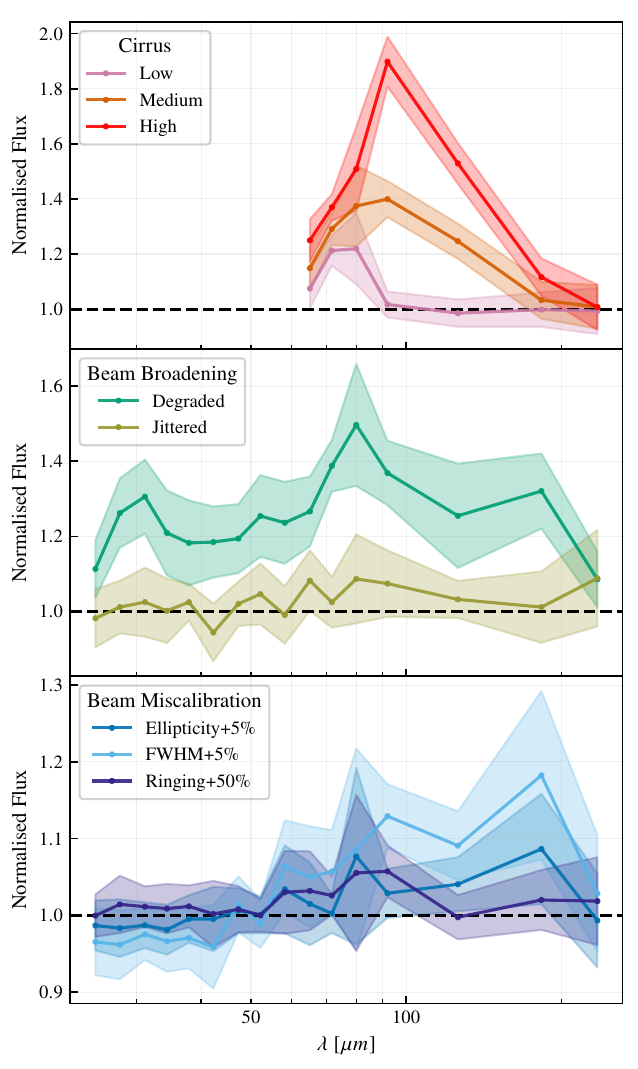}
    \caption{
    Impact of various systematics on the modelling accuracy of \xid across the PRIMAger bands.
    \textbf{Top:} Limiting flux achieved by \xid in the presence of simulated galactic cirrus, normalised by the limiting flux of our Fiducial run (dashed black line).
    We consider cirrus of varying intensities (0.5, 1.5, and 2.5~\mjysr) which are representative of the extragalactic fields PRIMAger is likely to observe.
    \textbf{Middle:} The impact of broadening the beam of the map on which \xid is run.
    This broadening could result from a smaller telescope aperture or some worsening of the optics (shown by the Degraded beams curve).
    It could also be the result of small positional offsets of target sources when stacking independent sky scans, leading to a blurring effect (shown by the Jittered curve).
    \textbf{Bottom:} The impact of supplying \xid with a beam which differs from the true beam of the map.
    These represent potential miscalibrations in the beam's ellipticity, FWHM, or ring features.
    Uncertainties are estimated via bootstrapping.
    }
    \label{fig:xid_variations}
\end{figure}

We consider low, medium, and high levels of cirrus with corresponding intensities of 0.5, 1.5, and 2.5~\mjysr (see Sections~\ref{sec:modelling_cirrus} and~\ref{sec:cirrus_contaminated_maps}).
We also look at two scenarios where the telescope beam is broader than assumed: degraded and jittered (Section~\ref{sec:non_nominal_beams}).
Finally, we investigate the impact of our \xid model beam differing from the beam used to create the map, looking at a miscalibration in the width, ellipticity, and strength of sidelobes (Section~\ref{sec:xid_beam_modelling}).
These runs are summarised in Table~\ref{tab:variation_runs}.
In all scenarios, we run \xid with the Euclid Wide positional prior, and a uniform flux prior (i.e. not \stepwise).
Figure~\ref{fig:xid_variations} shows the result of each of these runs.
For both the cirrus and beam-broadened scenarios, we provide \xid the same beam as was used to generate the maps, whereas we vary this for the miscalibration scenario.
Since cirrus intensity varies significantly across the field (Figure~\ref{fig:cirrus_structure}), we run a larger area of the simulated field for the cirrus-contaminated scenario.

The presence of cirrus at the aforementioned intensities, representative of those expected at the anticipated locations of the PRIMAger deep fields, increases our limiting fluxes by, at worst, a factor of $\sim$1.2, $\sim$1.4, and $\sim$1.9 respectively (Figure~\ref{fig:xid_variations}, upper panel).
In any of these scenarios we expect to be able to recover accurate SEDs out to PPI2 using \stepwise, for the two reference SEDs shown in Figure~\ref{fig:euclid_wide_limiting_flux}.
The channels that are most affected are consistently PHI2\_6, PPI1, and PPI2, with PPI3 and PPI4 being minimally affected. 
In the low cirrus scenario there is negligible impact on any of the PPI channels, whilst there is a noticeable impact on PHI2\_6.
The cirrus impact is a combination of the cirrus SED which peaks at $\sim$150~\um, between PPI2 and PPI3, the power spectrum of the cirrus fluctuations, and the local background estimator. 
Future work will explore their relative contributions.
However, our results show that \xid applied to PRIMAger will perform well in the presence of expected cirrus levels. 

In the beam-broadening scenario we compare the results of the broadened runs to those with our nominal beams (Figure~\ref{fig:xid_variations}, middle panel).
Due to how both runs are parametrised, we would expect shorter wavelength (i.e. narrower beams) to be disproportionately affected.
For the degraded beams, we find that our limiting flux increased by a factor of $\sim$1.25 on average, peaking at $\sim$1.5 in the PHI2\_6 channel.
With jittered beams, we instead find that the limiting flux increases by, at worst, $\sim$10\%, averaging only a 2\% increase over all channels. 

Finally, for the beam calibration runs, we compare them to a run where we give \xid the true map beam, the simplified beam described in Section~\ref{sec:non_nominal_beams}. 
We find small yet persistent improvements to our limiting flux for PHI1, although they are consistent with no change to within $1\sigma$ (Figure~\ref{fig:xid_variations}, lower panel).
Where we overestimate the beam width, we find that our limiting fluxes are, at worst, a factor of $\sim$1.18 higher at PPI3, and averaging $\sim$1.07 for PHI2 and PPI.
A slightly elliptical beam follows a similar relation, though roughly half as impactful, peaking at $\sim$1.09, again at PPI3, and averaging $\sim$1.03 for PHI2 and PPI.
Finally, even with a miscalibration of the beam of sidelobes 50\% stronger than nominal, our limiting fluxes are at worst a factor of $\sim$1.06 higher, averaging $\sim$1.02 for PHI2 and PPI.
This is as expected since the majority of the power of the beam lies within the central disc.
We expect the PRIMAger beams to be calibrated to higher precision than the variations considered here, so we are confident that \xid is resilient to realistically miscalibrated beam inputs.

\section{Discussion}
\label{sec:discussion}
Our results have shown that accurate flux measurements can be obtained well below the confusion limits of all the confusion-dominated PRIMAger channels.
This is achieved for sources detected in deep optical/NIR observations from both the \euclid Wide and Deep surveys and self-consistently with blindly detected sources.
Here we discuss the implications of our results for a specific science case as well as the limitations and requirements for future improvement of our work.

\subsection{IR-luminous galaxies at cosmic noon}
\label{sec:z2_sources}
In order to further evaluate the performance of our Fiducial run, we look at the specific science case of IR-luminous galaxies at cosmic noon ($z \sim$ 2).
These sources are of particular interest due to significant fractions of their star formation being dust-obscured and are observed at an epoch where cosmic star formation peaked.
Specifically, we look at sources within the simulated survey area which have $1.9 \leq z \leq 2.1$ and an LIR greater than the characteristic luminosity at this redshift, L$_{\text{IR}}^{*} = 10^{12.3}$L$_{\odot}$ \citep{Magnelli2013}.
As these sources have LIR$> 10^{12}$L$_{\odot}$, we refer to them as ultra-luminous IR galaxies (ULIRGs).
Given the low density of such sources, combined with the limited area being modelled through \xid ($\sim$0.2~\sqdeg), we run \stepwise on 22 additional non-contiguous tiles containing these ULIRGs, in order to have a sufficient number of sources for robust analysis.

In Figure~\ref{fig:euclid_wide_limiting_flux}, we show the model SEDs of two such sources from \cite{Kirkpatrick2015}, both of which have an L$_{IR}^{*}$ luminosity at $z=2$ but with differing AGN fractions, relative to the limiting fluxes reached by our three main runs.
A key science case of PRIMA will be to recover the SEDs of such sources in order to determine their physical parameters.
\cite{Bisigello2024} and \cite{Boquien2025} demonstrated that this is achievable if accurate flux measurements are obtained up to and including the PPI2 channel.
As shown in Figure~\ref{fig:euclid_wide_limiting_flux}, this is accomplished by our Fiducial as well as our Deep run since the limiting flux curves are well below the flux of these two model SEDs.
For the Blind run, however, the limiting flux lies just slightly above the two SEDs in PPI2, implying that their photometry will be recovered at a significance slightly lower than $5\sigma$ in that scenario. 

\begin{figure}
    \centering
    \includegraphics[width=\linewidth]{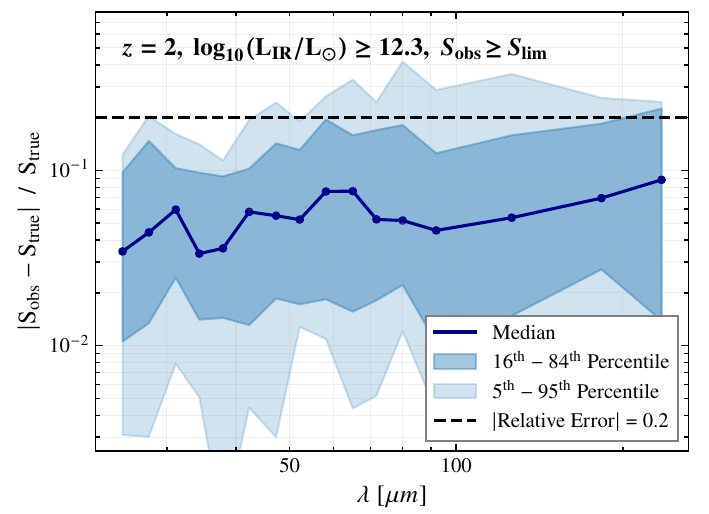}
    \caption{
    Distribution of the absolute relative errors between the observed and true fluxes from the Fiducial run for sources at z=2 with LIR $\geq 10^{12.3}$ L$_{\odot}$ and an observed flux above the limiting flux.
    }
    \label{fig:z2_rel_errs}
\end{figure}

Similarly to Figure~\ref{fig:euclid_wide_rel_errs}, we show in Figure~\ref{fig:z2_rel_errs} the distribution of the absolute relative errors between the observed and true fluxes of these $z=2$ ULIRGs which have an observed flux greater than the limiting flux from our Fiducial run.
Due to the lower number of sources in this narrow $z$ slice, this distribution is naturally more noisy than that for the full Fiducial run.
However, it is evident that the observed fluxes for these sources are accurate, with the median absolute relative error being well below the 0.2 threshold, corresponding to a flux accuracy of 20\% of the true flux.
Moreover, the 84$^{\text{th}}$ percentile is below this threshold for all but the longest wavelength channel, PPI4. 

Additionally, the impact of PAH emission is evident in the median absolute relative error, which drops for the PHI1\_4 and PHI1\_5 channels.
At this redshift, these channels cover the 11.3 and 12.7~\um PAH spectral features, which in turn raise the flux of these sources relative to the point source sensitivity.
This elevated signal-to-noise is then naturally reflected in the improved flux modelling accuracy.

\begin{figure}
    \centering
    \includegraphics[width=\linewidth]{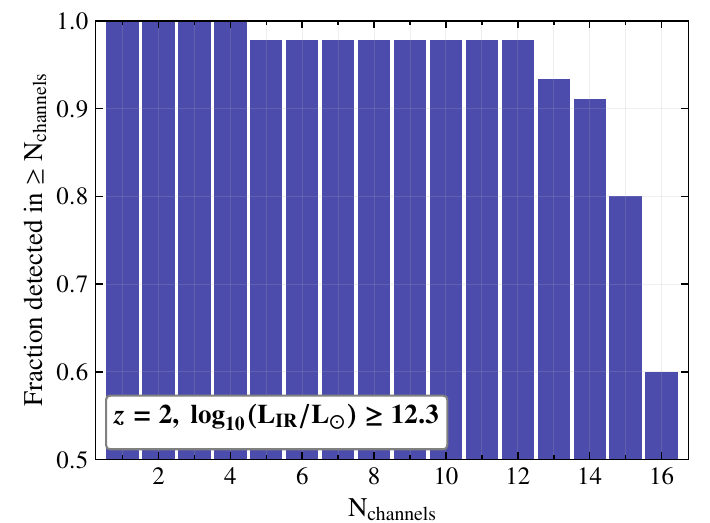}
    \caption{
    Cumulative histogram showing the fraction of $z=2$, LIR $\geq 10^{12.3}$ L$_{\odot}$ sources which are detected in a minimum number of channels.
    A source is considered detected if it has an observed flux greater than the limiting flux in a given channel.
    All of the $z=2$ ULIRGs in the simulated survey are detected in at least four channels, with 60\% having detections in all channels.
    Nearly all ($\sim$98\%) are detected in at least 12 of the PRIMAger channels, providing robust sampling of the SEDs of these scientifically interesting sources.
    }
    \label{fig:z2_n_bands_detected}
\end{figure}

If we again only consider sources with an observed flux above the limiting flux from our Fiducial run as being detected within a given band, we can determine the completeness of the $z = 2$ ULIRGs.
In Figure~\ref{fig:z2_n_bands_detected}, we show the cumulative fraction of these sources which are detected in a given number of the representative PRIMAger channels.
As shown, all of these sources are detected in at least four channels and $\sim$98\% of them have detections in at least 12 channels.
Moreover, 60\% of the $z = 2$ ULIRGs within the simulated survey area have detections in all channels.
As such, our Fiducial run is capable of providing robust sampling of the SEDs of these sources across the FIR wavelength range probed by PRIMAger. 

\begin{figure*}
    \centering
    \includegraphics[width=\linewidth]{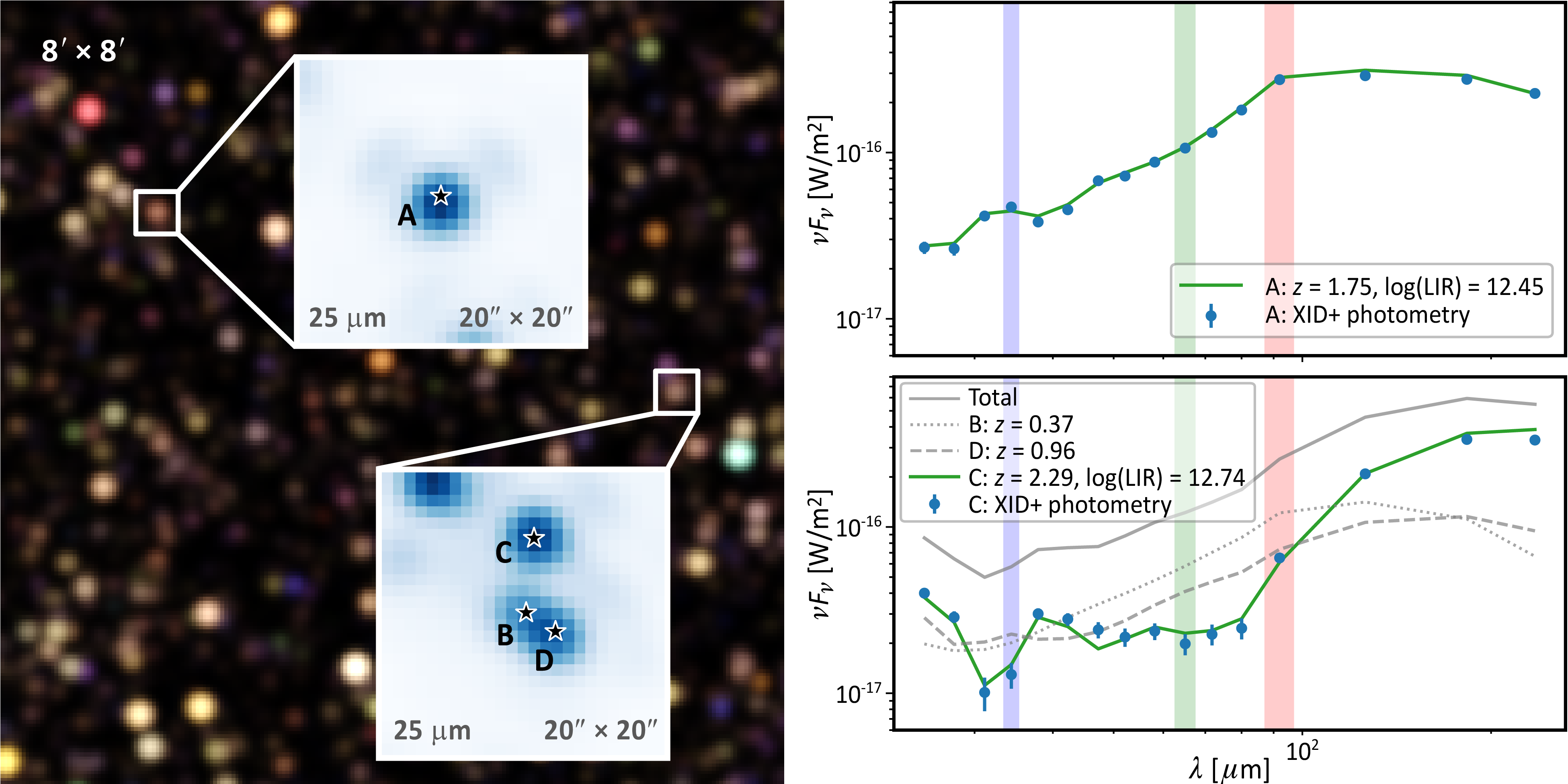}
    \caption{
    Demonstration of the deblending capabilities of \stepwise.
    \textbf{Left:} RGB cutout of the simulated PRIMAger map where blue, green, and red correspond to the PHI1\_4, PHI2\_4, and PPI1 channels respectively, with the resolution of each band degraded to that of PPI1.
    Also shown are insets of two regions of the PHI1\_1 map which contain an IR-luminous source with a redshift near cosmic noon.
    The top inset shows one of these sources (source A) which is relatively isolated, whilst the bottom shows another (source C) which has two close neighbours in the projected plane, but are at lower redshifts (sources B and D).
    As can be seen in the RGB image, sources B, C, and D are blended together at the resolution of PPI\_1.
    \textbf{Right:} Recovered SEDs of the cosmic noon, IR-luminous sources A and C from \stepwise photometry relative to their true SEDs from the SIDES simulation.}
    \label{fig:xid_demonstration_figure}
\end{figure*}

This is further demonstrated in Figure~\ref{fig:xid_demonstration_figure}, which highlights the capability of \stepwise in extracting accurate flux measurements for these sources from confusion-dominated maps.
We show the measured photometry from our Fiducial run relative to the true SEDs for two ULIRGs at cosmic noon.
The first, denoted as source A, is an isolated source relative to the size of the beam at PPI1 (which the resolution of the RGB image has been degraded to).
As such, the measured photometry for this source is within 7\% of the true flux at all of the PRIMAger channels.
The second source (denoted as source C), however, has two close neighbours which reside at lower redshifts, as shown by the bottom zoom inset of the RGB image in Figure~\ref{fig:xid_demonstration_figure}, depicting the PHI1\_1 intensity map at the location of these sources.
At the angular resolution of PPI1, these three sources are completely blended together.
As such, they significantly benefit from the utilisation of flux information obtained from the higher resolution maps in which they are not blended (i.e. via \stepwise).
This leads to all the recovered fluxes of source C being within 14\% of the true flux for all channels except for PHI2\_1, where the measured flux is $\sim$30\% off the true value.

\subsection{Future Work}
\label{sec:future_work}

Future work will look to further develop the simulated PRIMAger map generation as well as the modelling within \xid.
For the former, there are several aspects which can be improved.
First, our model of the extragalactic sky, the SIDES simulation, does not include AGN. Recently, however, \cite{Vidal2025} presented updates to the SIDES simulation which include modified star-forming and starburst SED templates as well as quiescent and AGN templates.
Second, including simulated zodiacal emission in addition to galactic cirrus would further ensure that the performance of \stepwise is robust to foreground contamination.
Third, realistic simulated instrumental noise will need to be incorporated which includes, for example, the spatial correlations from the telescope beam, 1/f noise, and effects from scanning strategies.
These are currently being developed as part of a realistic simulated map-maker for PRIMAger and will therefore be used in any future work testing flux recovery with \xid.

With regards to \xid, extending the modelling of the noise parameters to include the covariance from confusion noise will significantly improve the overall flux modelling, particularly in terms of the spread of the source flux posteriors (and therefore the quoted errors on the photometry).
Additionally, we will look to improve the modelling of the background and thereby resolve the underestimation of faint source fluxes, as described in Appendix~\ref{sec:appx_xid_biases}.

It is also possible to incorporate even tighter prior flux constraints within the \stepwise framework by utilising all the spectral filters within the two PHI bands.
In this work, we only considered using a combination of the available PHI filters.
Using all of them, however, allows for finer steps between wavelengths and thereby making the prior flux constraint obtained from one filter more accurate for the next filter.
An extension to \stepwise could also possibly fit the SEDs of the sources by constraining against the maps from all channels simultaneously.
This would be more expensive in terms of computer time as the number of free parameters being fit simultaneously scales with the number of channels included.
To do so, it would either require a smaller map tile size (and therefore fewer sources being modelled simultaneously) or employing GPUs.
Future work is thus needed to determine if the improved flux modelling and galaxy parameter constraints outweigh the increase in model complexity and computer time.

It is worth noting that other techniques for obtaining fluxes below the confusion limit are also possible, such as utilising auto-encoders and deep learning methods to super-resolve the confusion-dominated maps. \cite{Lauritsen21} and \cite{Koopmans25} demonstrated such techniques for \herschel/SPIRE 500~$\mu$m maps. A future paper (Koopmans et al., submitted) will explore deep learning based super-resolution methods in the context of PRIMAger.
\section{Conclusions}
\label{sec:conclusion}
In this work, we have demonstrated that accurate flux measurements can be obtained well below the confusion limits for all confusion-dominated PRIMAger channels.
By leveraging PRIMAger's hyperspectral capabilities, informative flux priors can be constructed by stepping through the spectral channels sequentially.
For \euclid-based positional prior catalogues, we have shown that the short-wavelength, higher angular resolution, PRIMAger data can be used to successfully deblend sources below the confusion limits in the longest wavelength PRIMAger maps.

We have tested these methods on mock data that simulate a deep 1500~hr~deg$^{-2}$ imaging survey with PRIMAger, from 25--235~\um.
These have been updated from \citetalias{Donnellan24} to reflect the latest PRIMAger instrument specifications.
As such, they have been generated using a sky, observatory, and instrument characteristics that provides maps with realistic instrumental and confusion noise. 

For our Fiducial run, we have shown that for a positional prior catalogue with the anticipated sensitivity of the \euclid Wide Field survey, our \stepwise methodology recovers accurate photometry of source fluxes (to within 20\%) which are a factor of 1.3--3.5 below the confusion limits between 45--84~\um.
Further gains were achieved for the most confused maps in the PPI channels (92--235~\um), with accurate flux measurements obtained at factors of $\sim$2--5 below the confusion limit.

For the Euclid Deep positional prior catalogue, we found that weak prior flux information is required at PHI1\_1 (25~\um) in addition to our \stepwise model in order to accurately recover fluxes down to the point source sensitivities of the PHI1 channels.
By doing so, we showed that even further gains below the confusion limits are achieved relative to our Fiducial run for all confusion-dominated channels, with up to a factor $\sim$1.4 deeper limiting flux at PPI1 (92~\um). 

By conducting blind source detection through matched-filtering of the PRIMAger maps, we also showed that accurate flux measurements from \xid below the confusion limits are currently achievable using data from PRIMAger in a self-contained way.
We have therefore demonstrated that imaging data from PRIMAger will not be limited by n\"aive, classical confusion noise if deblending with \xid is employed, with further gains possible from \euclid-like positional priors and \stepwise. 

From these results, we showed that accurate flux measurements were obtained with high spectral sampling from PRIMAger for $z = 2$ IR-luminous sources.
For those with an IR luminosity greater than the characteristic luminosity at $z = 2$, 60\% are detected in all 16 PRIMAger channels.
Moreover, nearly all ($\sim$98\%) are detected in at least 12 of the channels, thereby providing robust sampling of the FIR-regime of the SEDs of these dust-obscured, star-forming galaxies residing at the epoch of peak star-formation. 

We demonstrated that these results are robust to various systematic effects such as galactic cirrus, degraded optics, and various miscalibration effects.

\section*{Acknowledgements}
We thank David Leisawitz and Alexander Kashlinsky (NASA Goddard) for helpful discussions on the topic of modelling cirrus, and Darren Dowell (NASA JPL) for providing details and discussion on the beams.
We also thank John Arballo (NASA JPL) for help with the formatting of Figure~\ref{fig:xid_demonstration_figure}.
Finally, we would like to thank the anonymous referee for their helpful comments and feedback.
JMSD, BP, and SJO acknowledge funding from the UK Space Agency under grant UKRI3191.
JMSD was supported by the Science and Technology Facilities Council, grant number ST/W006839/1, through the DISCnet Centre for Doctoral Training.
AB, AP, and JDTS acknowledge funding for the PRIMA Phase A study from NASA.


\section*{Software}
This work made use of the following software packages:
\texttt{Astropy} \citep{astropy},
\texttt{IPython} \citep{IPython},
\texttt{JAX} \citep{jax},
\texttt{Matplotlib} \citep{matplotlib},
\texttt{NumPy} \citep{numpy},
\texttt{NumPyro} \citep{Numpyro2,Numpyro1},
\texttt{pandas} \citep{pandas},
\texttt{Photutils} \citep{Photutils3.0.0},
\texttt{SciPy} \citep{scipy}.
\\
\section*{Author Contributions}
The contributions of the authors using the Contributor Roles Taxonomy (CRediT)\footnote{\url{https://credit.niso.org/}} were as follows.
{\bf James Donnellan, Borja Pautasso}: Investigation, Methodology, Software, Validation, Writing - original draft, Writing - review \& editing,
{\bf Seb Oliver}: Conceptualization, Funding Acquisition, Project Administration, Supervision, Writing - Review \& Editing,
{\bf Matthieu B\'ethermin, Stephen Serjeant, Willem Jellema}: Software, Writing - Review \& Editing,
{\bf Longji Bing}: Investigation, Methodology, Software, Validation,
{\bf Alberto Bolatto, Laure Ciesla, Alex Pope, JD Smith}: Conceptualization, Writing - Review \& Editing,
{\bf Dennis Koopmans, Lingyu Wang}: Writing - Review \& Editing.

\section*{Data Availability Statement}
The data for this paper is available through Sussex Figshare:
\href{https:doi.org/10.25377/sussex.33123632}{https:doi.org/10.25377/sussex.33123632}.



\bibliographystyle{mnras}
\bibliography{main}

\appendix
\makeatletter
\renewcommand{\theHfigure}{appendix.\thesection.\arabic{figure}}
\renewcommand{\theHtable}{appendix.\thesection.\arabic{table}}
\makeatother
\section{Cirrus structure}
\label{sec:appx_cirrus}

We use the fractional Brownian motion (fBm) realisations of cirrus rather than existing observations to isolate the impact of cirrus emission, avoiding additional components often present in real maps (e.g. CIB, zodiacal light, map-making artifacts).
Furthermore, existing cirrus observations and commonly used tracers are typically observed at lower angular resolutions than PRIMAger, whether in the FIR, e.g. IRIS \citep{IRIS}, or radio, e.g. neutral hydrogen surveys such as HI4PI \citep{HI4PI}).
Finally, the fBm realisations allow us to investigate the impact of cirrus over a broad range of intensities, rather than being limited to the range observed by existing data.

The fBm process is characterised by one parameter, the Hurst exponent ($H$), which determines the correlation between successive changes.
When $H$ is 0.5 the field is uncorrelated, whereas values above or below this indicate a positive or negative correlation respectively.
For our work, we use $H$ = 0.7.

\begin{figure}
    \centering
    \includegraphics[width=\linewidth]{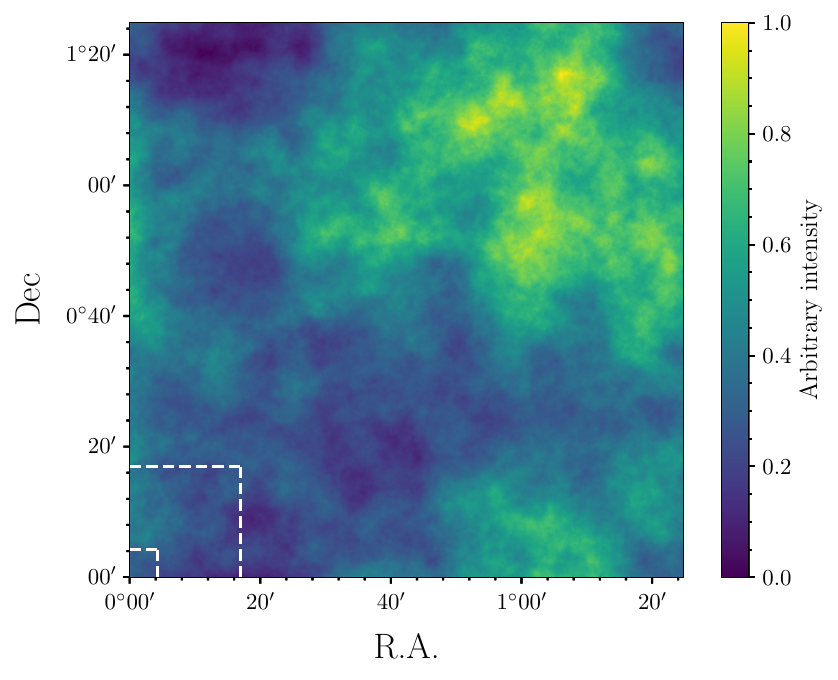}
    \caption{
    Structure of simulated cirrus over the entire 2~\sqdeg produced by the fBm process.
    The map is in arbitrary units.
    The small and large bounded areas correspond to the cutouts presented in Figures~\ref{fig:noisy_cutouts} and \ref{fig:cirrus_cutouts} respectively.
    }
    \label{fig:cirrus_structure}
\end{figure}

We first create the structure map with map pixels of 0\farcs1, and downscale to the PRIMAger pixel sizes.
After convolving the map with the beam, we measure the RMS via a double aperture chopping method as described in \cite{Gautier92}.
We normalise our maps to get the correct RMS, as predicted by \cite{Gautier92} and \cite{MD07}.
Although this will give us the correct variance at the scale of aperture, at other scales, both larger and smaller, this is likely to not be the case.
With the \cite{MD07} correction, the emission intensity scales linearly with \Ihundred.
Finally, we scale from the reference 100~\um to the desired wavelength by utilising the interstellar dust SED from \cite{Zubko04}.
The output of the fBm process is shown in Figure~\ref{fig:cirrus_structure}.

Before running \xid on the cirrus-contaminated maps, we first perform a local background subtraction process.
It estimates the background by first applying a 3$\sigma$ clipped median filter across the map, and a boxcar filter afterwards to smooth it out.

Future work will investigate the variance of the structure map at different spatial scales, alternative background subtraction algorithms, and the use of higher resolution imaging, such as the \textit{WISE} 12~$\mu$m cirrus maps \citep{Meisner_2014} or future \textit{SPHEREx} observations \citep{SPHEREx}.

\section{PRIMAger beams}
\label{sec:appx_beams}
\begin{figure*}
    \centering
    \includegraphics[width=\linewidth]{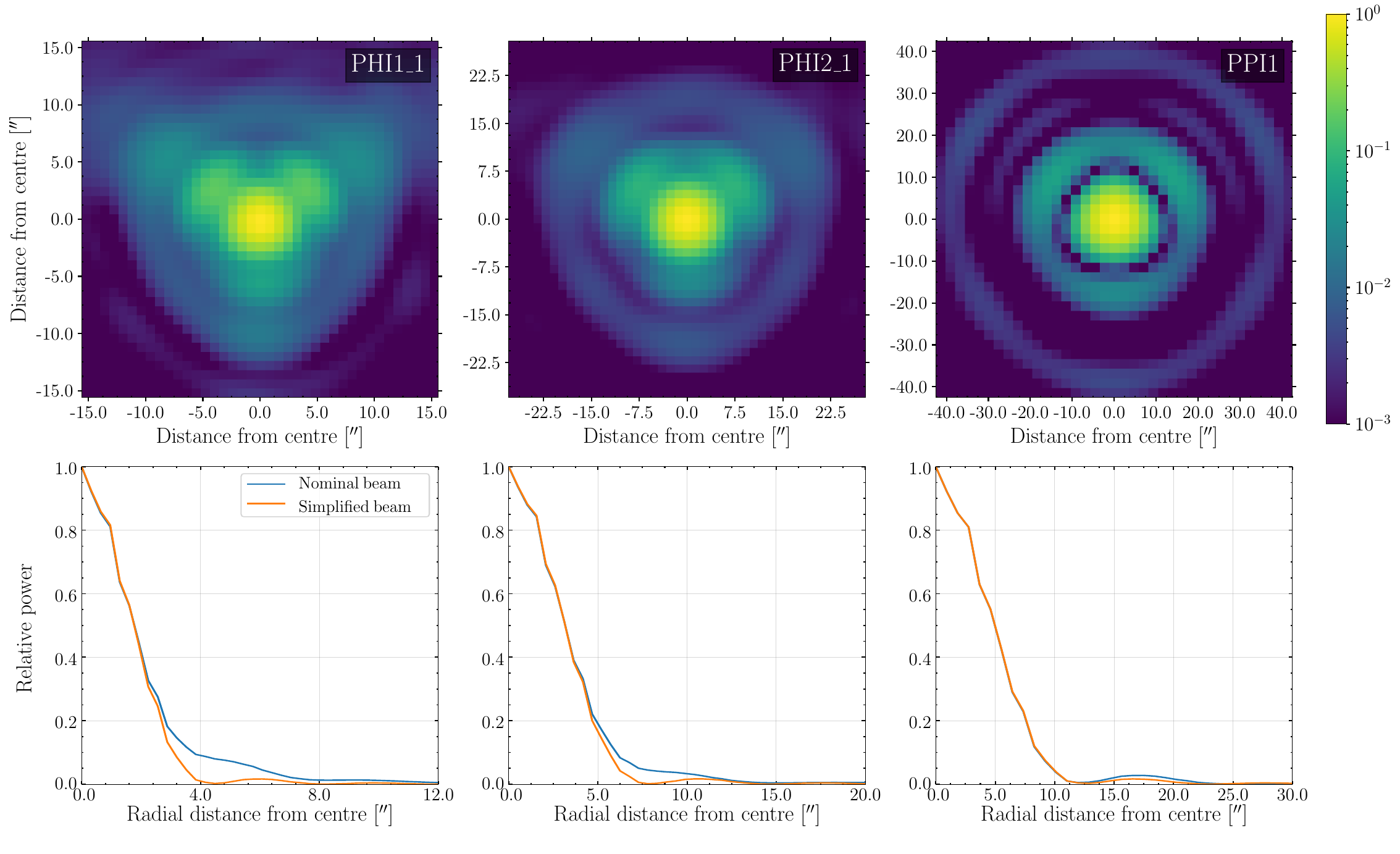}
    \caption{
    2D logarithmic view of beam PSF (top) and radial profile (bottom) for PHI1\_1, PHI2\_1, and PPI1.
    \textbf{Top:} 2D logarithmic view of the beam PSF for PHI1\_1, PHI2\_1, and PPI1.
    The various optical features present are apparent.
    \textbf{Bottom:} Radial profile of the nominal and simplified beam (Section~\ref{sec:non_nominal_beams}) after accounting for map pixelisation.
    }
    \label{fig:beams_2d}
\end{figure*}

Figure~\ref{fig:beams_2d} shows the 2D and radial profile of three representative PRIMAger channels: PHI1\_1, PHI2\_1, and PPI1.
The logarithmic view in the top panel shows the various effects introduced by including more realistic optical simulations.
The bottom panel shows the radial profile of both the nominal and simplified beams (Section~\ref{sec:non_nominal_beams}) after accounting for the map pixels.

\section{XID+ biases}
\label{sec:appx_xid_biases}

\begin{figure*}[h]
    \centering
    \includegraphics[width=\linewidth]{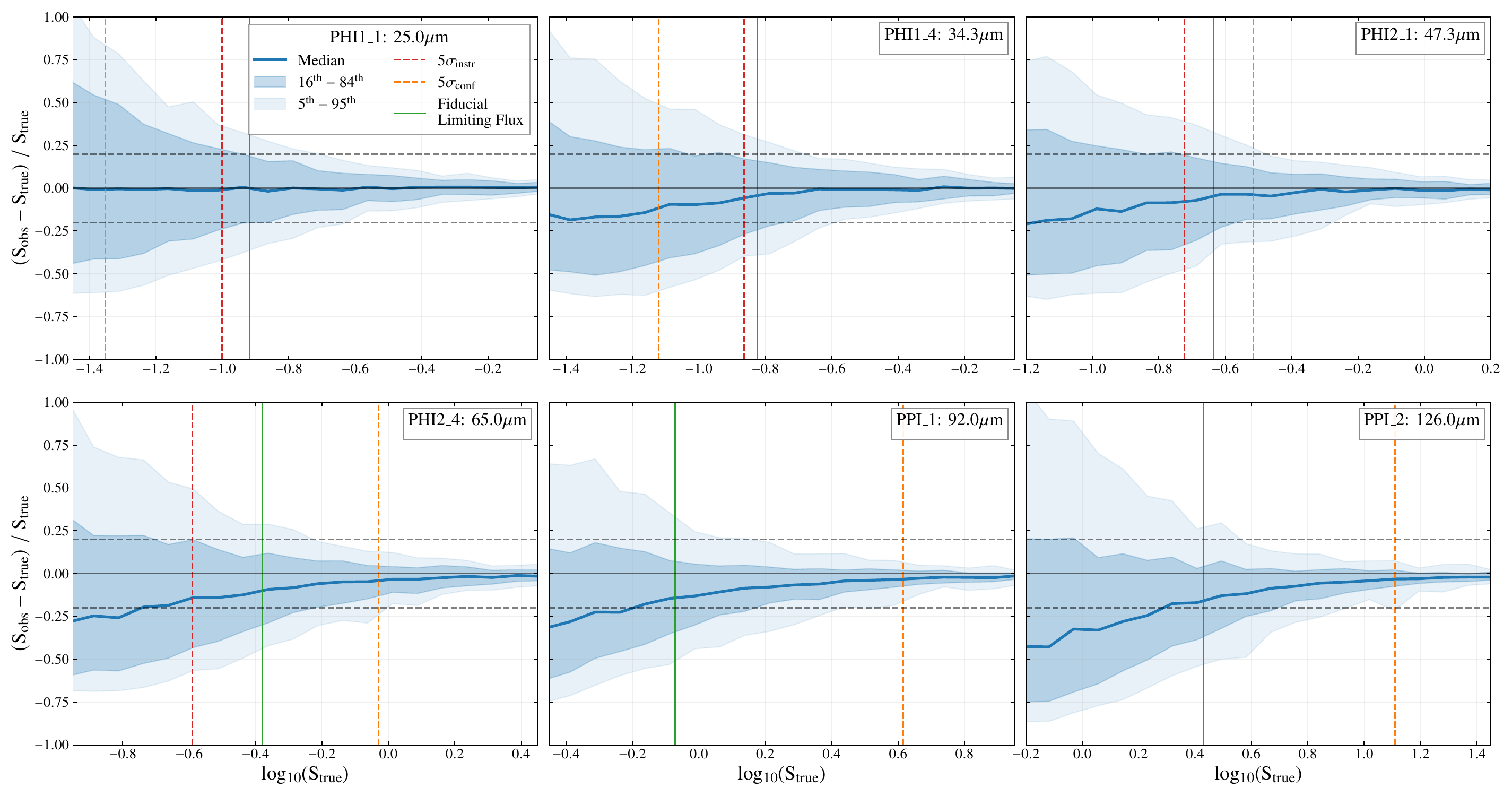}
    \caption{
    Distribution of relative errors between the observed and true fluxes from our Fiducial run as a function of the true flux of the sources for a selection of the PRIMAger bands.
    These highlight a bias from \xid to underestimate the flux, which worsens for fainter fluxes and longer wavelength.
    }
    \label{fig:appx_rel_errs_vs_true}
\end{figure*}

Figure~\ref{fig:appx_rel_errs_vs_true} looks at the distribution of relative errors between the observed and true fluxes from our Fiducial run as a function of the true flux of the sources.
These highlight a systematic underestimation in flux from \xid which becomes more pronounced for fainter sources and as the observed wavelength increases.
This arises due to a degeneracy between the background parameter and the residual confusion noise parameter within the modelling of \xid (Equation~\ref{eq:xid_eq}).
A solution with an elevated background value can be found either when bright sources are missing in the prior catalogue, or from bright sources being cut off at the edge of a tile during the tiling procedure.
Subsequently, extra flux is present in the map which can't be accounted for by the known sources in the model, as these may be too far from the regions of this unaccounted-for flux (as per the PRF).
This forces the background parameter to increase.
Since this background is also assumed to be constant across the tile, then the background will be too high in other areas and thus lead to a reduction in the fluxes of the known sources.

If this background parameter, which is a free parameter, is instead set to a constant value determined empirically to be the correct value, then this underestimation in the fluxes disappears.
Additionally, the residual confusion noise parameter increases to compensate.
Future work will investigate applying a non-uniform background model in order to resolve this issue. 

\begin{figure*}[h]
    \centering
    \includegraphics[width=\linewidth]{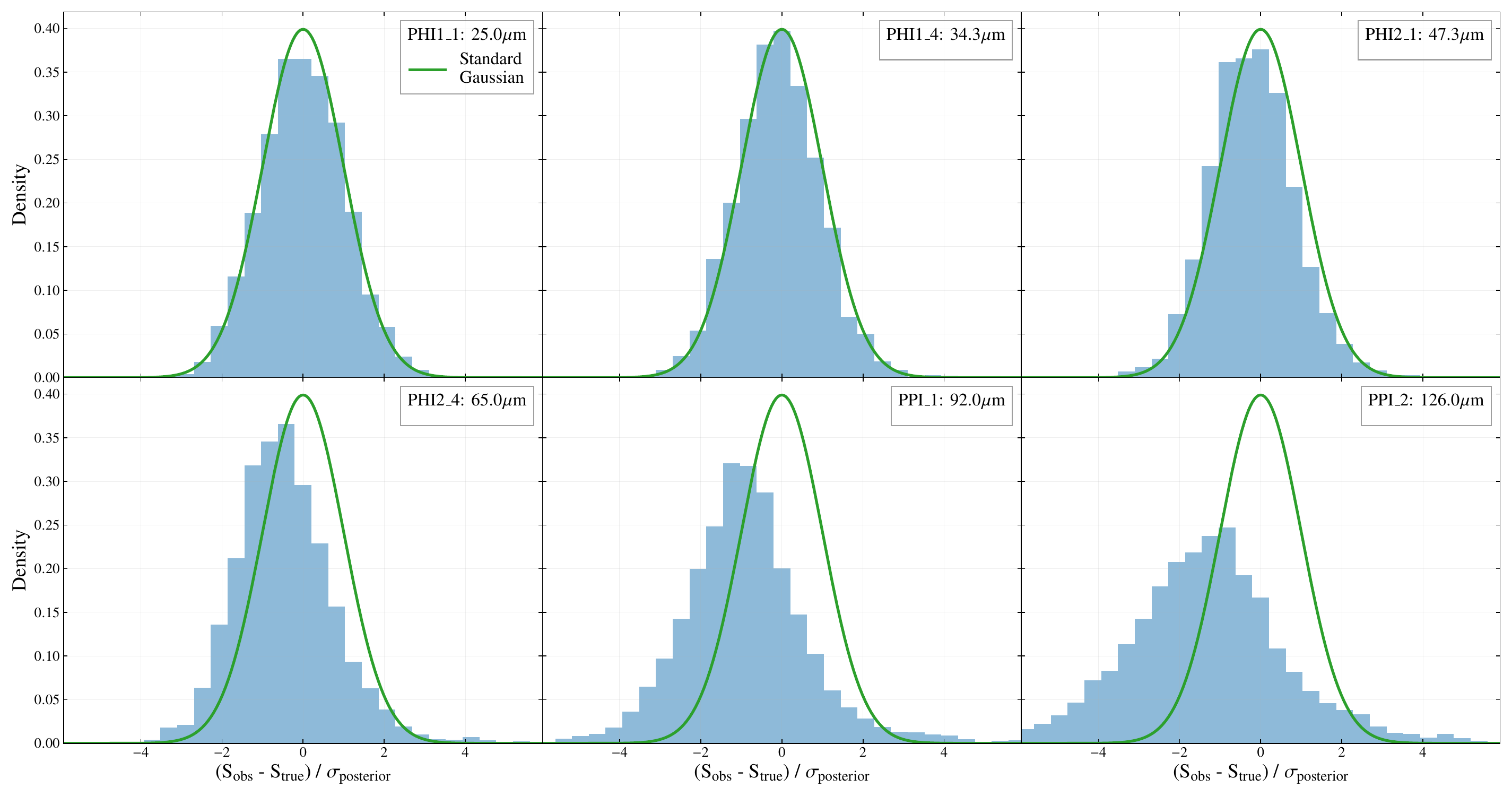}
    \caption{
    Histograms of the differences between the observed and true fluxes from our Fiducial run, normalised by the standard deviation of the individual source flux posteriors.
    The histograms only show sources which have an observed flux above the limiting flux of the Fiducial run in the given band.
    These are compared to a standard normal distribution ($\mu=0, \sigma=1$).
    For the longer wavelength channels, the histograms become less normally distributed and show that the spread in the source flux posteriors are underestimated.
    The bias to underestimate the flux (as shown in Figure~\ref{fig:appx_rel_errs_vs_true}) is also evident.
    }
    \label{fig:error_gaussianity}
\end{figure*}

We also look at the accuracy of the error bars on the fluxes measured by \xid in our Fiducial run.
These are taken as the maximum between the 16$^{th}$-50$^{th}$ percentile and the 50$^{th}$-84$^{th}$ percentile of the posterior samples for a given source from \xid fit.
This gives the 1$\sigma$ error in the measured flux of a source, which is taken as the median of the samples.
The accuracy of these errors can be checked by seeing how far away the true flux value of the source (S$_{\text{true}}$) is from the measured flux (S$_{\text{obs}}$), normalised by the error in the flux measurement ($\sigma_{\text{posterior}}$).
Ideally, this would correspond to the standard Gaussian distribution, whereby S$_{\text{true}}$ will be within 1$\sigma_{\text{posterior}}$ of S$_{\text{obs}}$ for 68\% of the sources. 

Figure~\ref{fig:error_gaussianity} shows the histograms of these values for sources within our Fiducial run which have an observed flux above the limiting flux in a given channel.
For the PHI1 channels, these histograms do indeed match the expected standard Gaussian.
However, for the PHI2 and PPI channels, they move away from the standard Gaussian and become ever more biased towards negative values.
This is due to the bias in \xid to underestimate the flux, as discussed above and shown in Figure~\ref{fig:appx_rel_errs_vs_true}.
What is also evident is the increase in the spread of the histograms relative to the standard Gaussian for these channels.
This indicates that the errors on the measured fluxes (i.e. the spread in the flux posteriors) become increasingly underestimated as the wavelength increases.
This is due to the residual confusion noise parameter within the \xid model (Equation~\ref{eq:xid_eq}) not taking into account the correlations between pixels.
Future work will look to account for this covariance within the modelling.

\section{XID+stepwise Extracted SEDs}
\label{sec:appx_xid_seds}

\begin{figure*}[h]
    \centering
    \includegraphics[width=\linewidth]{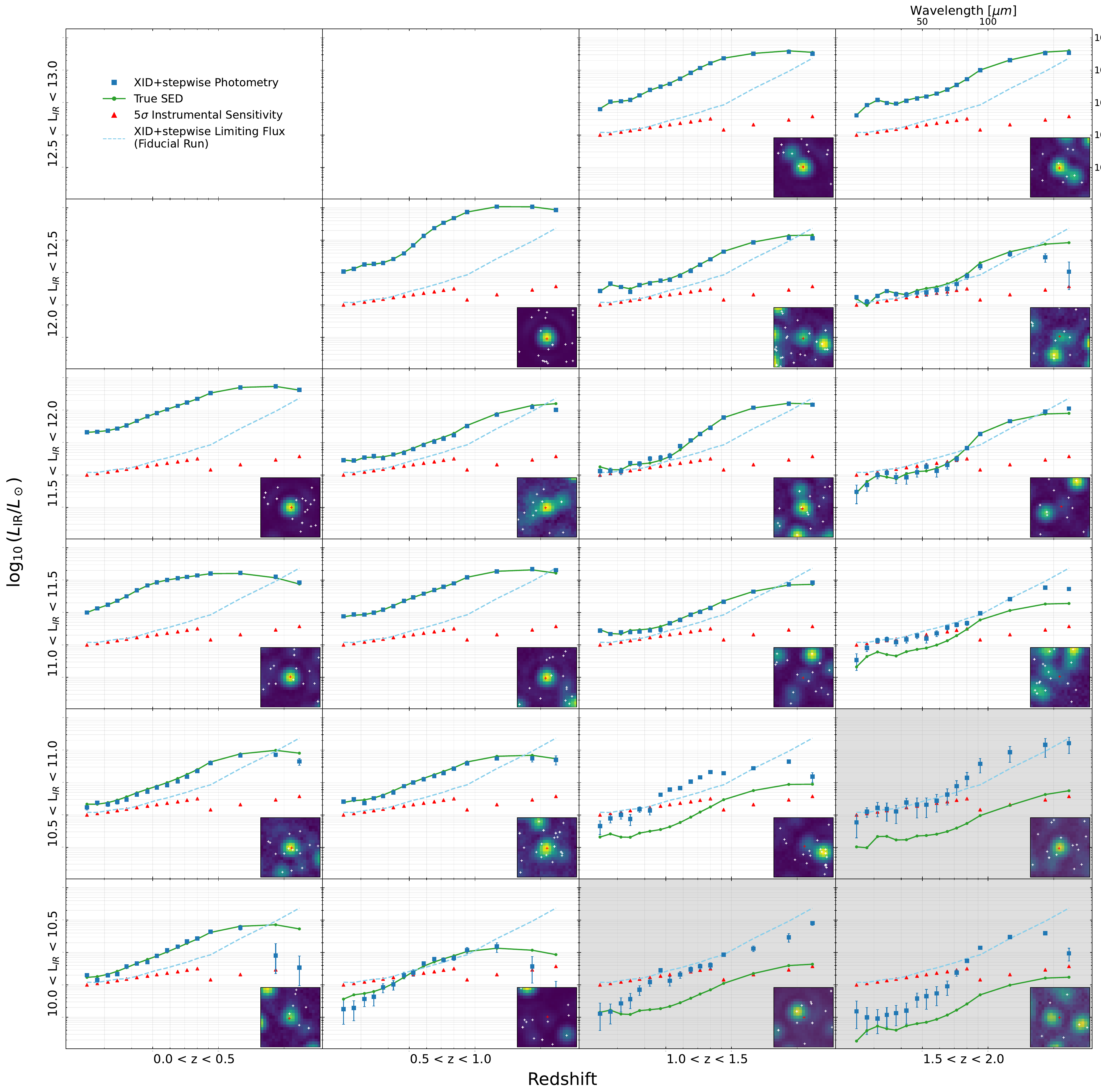}
    \caption{
    Extracted SEDs for randomly selected sources within redshift-LIR bins. Sources within a bin are selected if the observed flux is above the limiting flux in at least two channels.
    For each bin, the \stepwise photometry from our Fiducial run for all 16 PRIMAger channels are shown by the blue squares.
    The quoted photometry values are given by the median of the source flux posterior from \xid, with the 16$^{\text{th}}$ and the 84$^{\text{th}}$ percentiles shown by the error bars.
    The true SED of each source from the SIDES light-cone catalogue is shown by the solid green curve.
    For reference, the limiting flux values from our Fiducial run are shown by the dashed blue curve.
    Additionally, the 5$\sigma$ instrumental point source sensitivities are shown by the red triangles.
    Also plotted in the bottom-right of each panel is a 50~$\times$~50\arcsec\ cutout of the simulated PPI1 map, centred on the source for which the extracted and true SEDs are shown (marked by the red cross).
    Neighbouring sources from the Euclid Wide prior catalogue have their positions marked by the white crosses.
    Empty panels are due to no sources from the Fiducial run being present within the given redshift-LIR bin.
    Grey-shaded panels contain no sources within the given bin which have a true flux above the instrumental sensitivity in at least half of the channels. 
    }
    \label{fig:extracted_seds}
\end{figure*}

Figure~\ref{fig:extracted_seds} shows a selection of SEDs extracted from our Fiducial run for sources with a range of redshifts and IR luminosities.
The sources shown are randomly selected within redshift-luminosity bins ranging from $0 < z < 2$ and $10 < \text{log}_{10}(\text{LIR} / \text{L}_{\odot}) < 13$ if they have an observed flux above the limiting flux in at least two channels.
The measured photometry from our Fiducial run is compared to the true values for each source to demonstrate how the accuracy and precision of the extracted SEDs change relative to the effective signal-to-noise of the source.
For sources at lower redshifts and higher IR luminosities (i.e. panels towards the top left of the figure), the flux of the source across all channels is higher relative to both the instrumental noise and the \xid limiting flux.
As expected, \stepwise performs exceptionally well in terms of extracted flux accuracy.
The performance then gradually degrades, as expected, for those at higher redshifts and lower LIR (i.e. sources in panels towards the bottom right of the figure).

\subsection{Pathological Failures}
\label{sec:appx_failures}

\begin{figure*}[htbp]
    \centering
    \includegraphics[width=\linewidth]{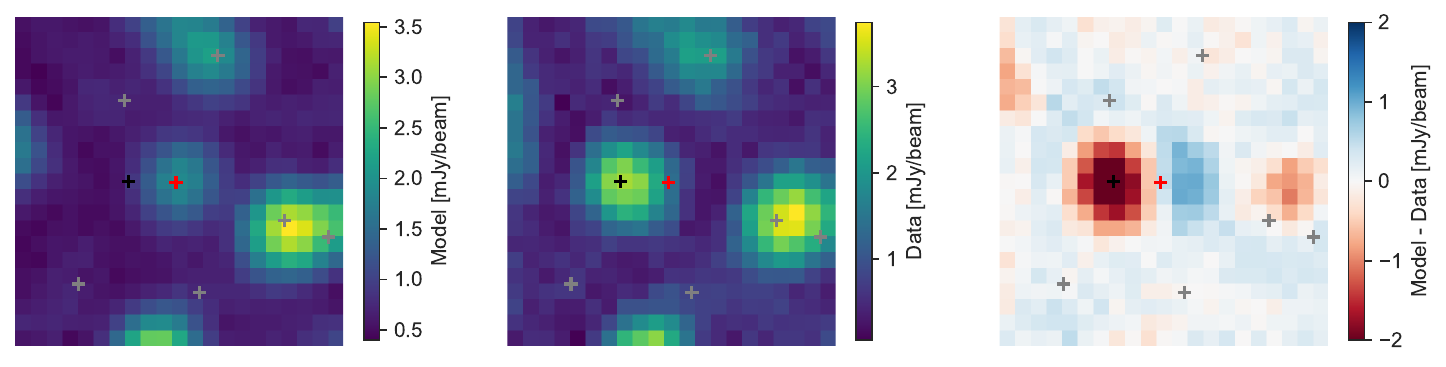}
    \caption{
    50~$\times$~50\arcsec\ cutouts of the Fiducial \stepwise modelled map for the PPI1 channel (left), the simulated data map (middle), and the residual map (right) centred on the source from the $1.5 < z < 2.0$ and $10 < \text{log}_{10}(\text{LIR} / \text{L}_{\odot}) < 10.5$ bin of Figure~\ref{fig:extracted_seds}.
    The \xid flux modelling is confidently incorrect for this source, which is marked by the red cross in the cutouts, due to flux boosting from a nearby bright source which is missing from the Euclid Wide prior catalogue.
    The position of this missing source is shown by the black cross, whilst all other prior catalogue source positions are marked by gray crosses.
    The residual map shows how the flux from the missing source leaks into the nearby sources which are known to \xid.
    It also demonstrates that these bright sources which are missing from our prior catalogue can still be detected.
    }
    \label{fig:missing_source_residual}
\end{figure*}

One notable source in Figure~\ref{fig:extracted_seds} is shown in the $1.5 < z < 2.0$ and $10.0 < \text{log}_{10}(\text{LIR} / \text{L}_{\odot}) < 10.5$ bin, whereby the extracted photometry is consistently boosted above the true flux values of the source.
The measured photometry is also confidently above the \xid limiting flux in the PPI1 and PPI2 channels, despite the true fluxes being well below this threshold.
The modelling for this source has therefore failed and is due to the flux boosting caused by a bright source which is missing from the Euclid Wide prior source position catalogue.
Most importantly, this is detectable within the residual between the \xid model map and the real map, as shown in Figure~\ref{fig:missing_source_residual}.
The residual map shows how the flux from the missing source leaks into the nearby sources which are known to \xid.
The missing source is still identifiable, however, due to pixels surrounding it having too little flux in the model compared to the data map.
It is worth noting that a second, albeit fainter, missing source is also identifiable in the residual map to the right of the target source.
Additionally, this methodology could also be utilised to identify extended sources if they have incorrectly been assumed to be point sources within \xid.

To identify other pathological failures, and to determine the rate at which they occur, we define the following failure criteria based upon the modelled and true fluxes.
The modelling of a source is considered a failure if the modelled flux is above the limiting flux in a given channel and for which it is confidently incorrect by a factor of $\geq 2$ with respect to the true flux of the source.
In order to be considered as confidently incorrect, the absolute difference between the modelled and true flux of the source must be more than 5$\sigma_{\text{posterior}}$.

With this, the number of sources with failed modelling in our Fiducial run slowly rises from 0.1\% of sources above the limiting flux in the PHI1\_1 channel, to 1.0\% and 2.0\% in the PHI2\_1 and PPI1 channels respectively.

The majority of these failures are due to nearby, bright sources which are missing from the prior source catalogue.
This mode of failure is, therefore, significantly reduced for the Deep run due to the Euclid Deep prior catalogue containing nearly every IR-bright source, as shown in Figure~\ref{fig:euclid_cats_percentage_wvs}.
For this run, no modelled sources meet this criteria in the PHI1\_1 channel due to the additional prior flux information provided in this first step.
The modelling failure rate then only increases to 0.05\% and 0.1\% of sources above the limiting flux in the PHI2\_1 and PPI1 channels respectively. 

\section{Computational Requirements}
\label{sec:appx_compute}

\begin{deluxetable}{llcccc}[h]
\tablecaption{Compute Statistics Summary}

\tablehead{
\colhead{Run} &
\colhead{Channel} &
\colhead{$N_\text{eff}$} &
\colhead{$\hat{R}$} &
\colhead{$t_i$} &
\colhead{$t_i/N_{\text{eff},i}$}
\\
\colhead{} &
\colhead{} &
\colhead{} &
\colhead{} &
\colhead{[CPU h]} &
\colhead{[CPU s]}
}

\startdata
Fiducial & PHI1\_1  & 4200  & 0.9999 & 0.38 & 0.32 \\
Fiducial & PHI2\_1  & 3970  & 0.9999 & 0.19 & 0.17 \\
Fiducial & PPI1     & 3780  & 1.0000 & 0.12 & 0.11 \vspace{5pt}\\

Deep     & PHI1\_1  & 3810  & 1.0000 & 1.63 & 1.56 \\
Deep     & PHI2\_1  & 3890  & 1.0000 & 0.61 & 0.55 \\
Deep     & PPI1     & 3750  & 1.0000 & 0.37 & 0.37 \vspace{5pt}\\

Blind    & PHI1\_1  & 4680  & 0.9997 & 0.23 & 0.17 \\
Blind    & PHI2\_1  & 3860  & 0.9999 & 0.08 & 0.07 \\
Blind    & PPI1     & 4040  & 0.9999 & 0.04 & 0.03 
\enddata
\tablecomments{
Summary compute statistics of our \xid runs, split by prior catalogue and channel.
For each, we report the effective sample size ($N_\text{eff}$), the potential scale reduction factor ($\hat{R}$), the runtime of an individual order 11 tile ($t_i$), and the cost per effective sample ($t_i/N_{\text{eff},i}$).
$N_\text{eff}$ and $\hat{R}$ are shown to 3 significant figures and 4 decimal places respectively, and give the median over all sources in all tiles.
$t_i$ and $t_i/N_{\text{eff},i}$ are shown to 2 decimal places, and give the median over all order 11 tiles, where $N_{\text{eff},i}$ is the median $N_\text{eff}$ for tile $i$.
Of the nine scenarios shown, only the Deep run contains any sources with an $\hat{R} > 1.01$.
One source, which has a nearby neighbour missing from the prior catalogue, appears in both PHI1\_1 and PHI2\_1, whilst two other sources, which are themselves neighbours of each other, appear in PPI1.
}
\label{tab:compute_table}
\end{deluxetable}

We parallelise the \xid framework using the HEALPix tiling scheme \citep{Healpix}, following the approach of \citet{Pearson2017}.
HEALPix splits the sky into equal area pixels, allowing large maps to be divided into independent regions that can be processed in parallel.

Throughout this work, we refer to HEALPix resolution by its order, where increasing the order subdivides a pixel into four smaller pixels (i.e. one order 7 pixel is composed of four order 8 pixels).

For each of our three prior catalogues, we first extract an order 7 cutout (covering $\sim$0.21~deg$^2$) from the full 2~deg$^2$ maps, containing 20,445, 54,270, and 8,605 sources for the Euclid Wide, Euclid Deep, and Blind catalogues respectively.
This region is then further subdivided into 256 order 11 tiles ($\sim$2.95~arcmin$^2$), on which \xid is run.

We choose order 11 tiles as a compromise between computational efficiency and modelling accuracy.
Larger tiles contain more sources and pixels, which provide stronger constraints on source and background parameters, but also increase the computational cost.

To mitigate edge effects, each order 11 tile is expanded by including a surrounding ring of order 13 tiles prior to modelling.
This padding ensures that sources outside the target region, but whose emission extends into the modelling area, are included in the fit.
The padded region is only used during modelling, and is discarded for the final results.

For each tile we run a NUTS sampler implemented with \texttt{NumPyro} to model the posterior distribution.
We run four independent MCMC chains per tile, each consisting of 1,000 warm-up and 1,000 regular samples.
For each tile we store the full prior and posterior objects, alongside summary statistics describing the inferred flux distribution for each source.
The posterior object allows for subsequent analysis of the posterior distribution, whereas the summary statistics are later merged to produce a master catalogue for each prior catalogue.

These runs were performed on the Artemis high performance computing cluster (HPC) within the University of Sussex.
Each tile was run as an independent job, with a single CPU core (AMD EPYC Milan) and 4 GB of RAM being allocated.
Due to the small area being modelled, all four chains are able to fit within a single core, allowing all to be run in parallel with \texttt{NumPyro}'s `vectorized' chain method.

In Table~\ref{tab:compute_table}, we report two MCMC diagnostics: the effective sample size ($N_\text{eff}$) and the potential scale reduction factor ($\hat{R}$); alongside two efficiency diagnostics: the runtime of an individual order 11 tile ($t_i$) and the cost per effective sample ($t_i/N_{\text{eff},i}$), split by prior catalogue and a subset of channels.
In all cases, the median values indicate the chains are well mixed ($\hat{R} < 1.01$) and give a large number of effective samples ($N_\text{eff} > 1000$).

We have demonstrated that \xid can be efficiently applied to large fields through the HEALPix tiling scheme.
This parallelisation would allow an arbitrarily large field to be run in parallel on an HPC, if resources permit.
Extrapolating from Table~\ref{tab:compute_table}, all 16 channels of any such field at a depth of our Euclid Deep catalogue could be run sequentially through \stepwise in $\sim$20~hours (wall time).

\end{document}